\documentclass[reprint, pra, twocolumn,  amsmath,   amssymb, 
  floatfix,fleqn]{revtex4-1}
\usepackage{eqnarray,graphics, graphicx, bm, natbib, color, xcolor}

\usepackage[capitalise]{cleveref}

\newcommand{\rem}[1]{}

\newcommand{\ud}{\mathrm{d}}

\newenvironment{psmallmatrix}
  {\left(\begin{smallmatrix}}
  {\end{smallmatrix}\right)}

\begin{document}

\author{Wijnand Broer}
\email{wbroer@gmail.com}
\affiliation{Division of Physics and Applied Physics, School of Physical and Mathematical Sciences, Nanyang Technological University, 21 Nanyang Link, Singapore 637371}

\author{John Yuh Han Liow}

\affiliation{Division of Physics and Applied Physics, School of Physical and Mathematical Sciences, Nanyang Technological University, 21 Nanyang Link, Singapore 637371}

\author{Bing-Sui Lu}
\email{bslu@ntu.edu.sg}
\affiliation{Division of Physics and Applied Physics, School of Physical and Mathematical Sciences, Nanyang Technological University, 21 Nanyang Link, Singapore 637371}

\title{Maxwell Eigenmode approach to the Casimir-Lifshitz Torque}

\date{\today}

\begin{abstract}

\noindent{More than forty years ago, Barash published a calculation of the full retarded Casimir-Lifshitz torque for planar {birefringent} media with arbitrary degrees of anisotropy. An independent theoretical confirmation has been lacking since. We report a systematic and transparent derivation of the torque between two media with both electric and magnetic birefringence. Our approach, based on an eigenmode decomposition of Maxwell's equations, generalizes Barash's result for electrically birefringent materials, 
{and can be generalized to}
 a wide range of anisotropic materials and finite thickness effects.}
 
\end{abstract}

\maketitle

\section{Introduction}

\noindent Casimir-Lifshitz forces~\cite{Casimir48} are dispersion interactions between macroscopic bodies that arise from quantum mechanical and  thermal fluctuations in the electromagnetic field. These forces, which can be considered a generalization of van der Waals forces to include finite light speed, depend on the electric and magnetic susceptibilities of the materials involved ~\cite{Lifshitz55,*Lifshitz61}  ({cf. also} Ref.~\cite{Woods2016Review} for a review).  Several decades  after its first theoretical prediction, measuring the Casimir force directly became technologically feasible~\cite{Lamoreaux97}. Since this interaction is mediated by virtual and thermal photons, the frequency of which cannot be controlled directly, the Casimir force is a broadband phenomenon. In particular, as a consequence of the fluctuation-dissipation theorem, the frequency ranges where the susceptibilities change significantly actually provide a dominant contribution to the Casimir force. 

From a fundamental viewpoint, the Casimir force plays a role in micron range gravitation experiments, and
 the search for deviations from Newtonian gravitation due to hypothetical new forces~\cite{George2015}. This fuels the desire to come to precise comparisons between theoretical predictions and experimental data~\cite{Almasi2015}. More practically, Casimir interactions affect the actuation dynamics of nano- and micro-mechanical systems, such as switches, cantilevers, and actuators at a sub-micrometer length scale~\cite{Serry1995,Chan2001,*ChanScience2001,Miri2008, Broer2015,Svetovoy2017, Klimchitskaya2018, Tajik2018PRE, Ahn2018}.

\begin{figure}[!htbp]
	\centering
		\includegraphics[width=0.48\textwidth]{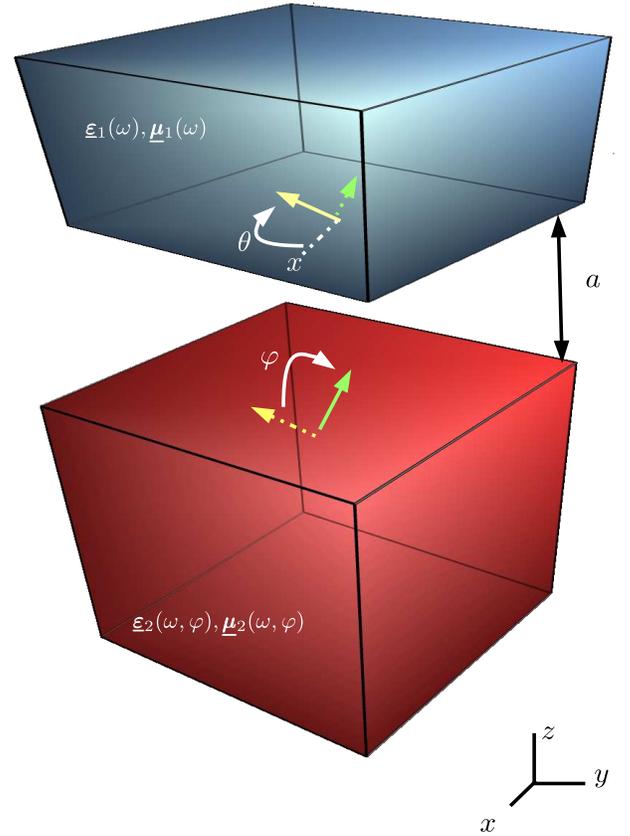}
	\caption{Two half spaces with both electric and magnetic birefringence. The optic axes of both media are indicated by the solid arrows. The dash-dotted arrows show the projection of the optic axis of the other medium. The $x$-$y$-plane refers to the laboratory's coordinate {system, which is identical for both surfaces.}}
	\label{fig:fig1}
\end{figure}

In order to calculate the Casimir potential, the Maxwell equations must be solved for the given geometry~\cite{Lifshitz55,*Lifshitz61}. Here we focus on the  case of planar media where this can be done analytically. Anisotropy in the plane of reflection creates a dielectric contrast in the azimuthal direction, which gives rise to a Casimir torque~\cite{Barash1978} ({see}~\cref{fig:fig1}). Several experimental setups to detect the torque have been proposed~\cite{Munday2005, Rodrigues2008, Guerout2015,XuZ2017}, but  only recently has this phenomenon been observed experimentally~\cite{Somers2018}.

An exact analytical description for two planar birefringent half spaces was derived more than forty years ago {by Yu. Barash~\cite{Barash1978}.} More recently, an alternative calculation has been presented~\cite{Philbin2008}, the result of which looks symbolically different. It has not been established that this result agrees analytically with that of Ref.~\cite{Barash1978}, though the authors claim to have verified it numerically. 

Recently, another formula  was proposed for the Casimir-Lifshitz torque {in appendix A of Ref. ~\cite{Somers2017},  where the result of of Ref. \cite{Lekner1991} is inserted into the Lifshitz formula. \cite{Lambrecht2006} However, this does not include a} supporting calculation that shows that it is equivalent to the result of Ref.~\cite{Barash1978}. Up to the present, a transparent and independent calculation which analytically verifies the result of Barash has still been lacking. 

The recent  development of the experimental observation of the Casimir torque~\cite{Somers2018} will instigate more investigations on this phenomenon, which necessitates a systematic and transparent formalism to describe it theoretically. The main obstacle is the failure of the usual decomposition into perpendicular $s$-polarized (or transverse electric) and $p$- polarized (transverse magnetic) modes. The reason for this failure is the fact that these are not the solutions of the Maxwell equations. Therefore it stands to reason to determine what these solutions actually are instead.  This can be done by formulating the Maxwell equations as an eigenvalue problem. The Maxwell eigenmode formalism is a well-established method in electrical and electronic engineering~\cite{Berreman1972, Yeh1979} that has been designed specifically to tackle the problem of scattering electromagnetic waves on anisotropic media. Note that the $s$- and $p$-mode decomposition actually does work if the plane of anisotropy is perpendicular to the plane of reflectance (cf., e.g., Ref. \cite{Grushin2011}). However, in such a case the Casimir torque vanishes.

Some limiting cases simplify the result of Ref.~\cite{Barash1978} considerably. {An example is} the limit of weak anisotropy, or more precisely, the limit of relatively small deviations from in-plane anisotropy{. Although a treatment of the full anisotropic case exists \cite{RosaPRA2008}, this is still a popular approximation~\cite{Esquivel2010,SomersPRL2017,Thiyam2018}.} This is valid for certain natural anisotropic materials such as calcite or quartz, but there is no reason to assume this must hold generally. Examples of materials that are not weakly anisotropic include, but are not limited to, to cuprate superconductors, \cite{Romanowsky2008} liquid crystals, \cite{Jakli2013} anisotropic metamaterials, \cite{Poddubny2013} and multiferroic materials \cite{Fiebig2016}. Another common simplifying assumption is the non-retarded limit of van der Waals forces~\cite{Bing2016,Bing2018}. Depending on the material(s), this approximation should work at separation distances of the order of 10 nm. However, even at such short distances this approximation can fail~\cite{SomersPRL2017}. Here, we would like to make the case that such approximations are unnecessary by presenting an exact and transparent generalization of the calculation in Ref.~\cite{Barash1978}. 

The approach that we present is subject to the same assumptions underlying the original Lifshitz theory. Firstly, it 
relies on a continuous medium approximation, which is valid for wavelengths larger than the interatomic distance. A second assumption is that the medium exhibits linear dielectric response. Indeed, conventional Lifshitz theory does not extend to media with nonlinear dielectric response behavior. Such an extension requires a non-trivial generalization of the fluctuation-dissipation theorem, ~\cite{Soo2016,*Soo2018}. which is based on linear response theory

\section{Maxwell eigenmodes}

In the case of birefringent media, one can distinguish the so-called `ordinary' and `extraordinary' waves. The former propagates as if the medium were isotropic, whereas the propagation of the latter depends on the medium's orientation~\cite{LandauLifshitz8}. 

Let the material slab be oriented in such a way that the anisotropic plane is facing the surface, which is defined as the $x$-$y$-plane in the laboratory's coordinate system. Furthermore, the magnetic and electric anisotropy axes are assumed to be identical. Hence the electric permittivity and the magnetic permeability are given by the following tensors:

\[\underline{\pmb{\varepsilon}}(\omega)=\begin{psmallmatrix}
	\varepsilon_{1x}\cos^2\theta+\varepsilon_{1y}\sin^2\theta&(\varepsilon_{1x}-\varepsilon_{1y})\sin\theta\cos\theta&0\\
	(\varepsilon_{1x}-\varepsilon_{1y})\sin\theta\cos\theta&\varepsilon_{1x}\sin^2\theta+\varepsilon_{1y}\cos^2\theta&0\\
	0&0&\varepsilon_{1y}
\end{psmallmatrix}
\]
\[\underline{\pmb{\mu}}(\omega)=\begin{psmallmatrix}
	\mu_{1x}\cos^2\theta+\mu_{1y}\sin^2\theta&(\mu_{1x}-\mu_{1y})\sin\theta\cos\theta&0\\
	(\mu_{1x}-\mu_{1y})\sin\theta\cos\theta&\mu_{1x}\sin^2\theta+\mu_{1y}\cos^2\theta&0\\
	0&0&\mu_{1y}
\end{psmallmatrix}
\]
where $\theta$ denotes the angle between material's  optic axis  and the $x$-axis of laboratory's coordinate system ({cf.}~\cref{fig:fig1}). It must be stressed that the entries of both $\underline{\pmb{\varepsilon}}$ and $\underline{\pmb{\mu}}$ depend on frequency, but that the argument will be suppressed from now on.

{The {vectorial Maxwell} equations in Fourier space are given by}

\begin{equation}
\mathbf{k}\times\mathbf{E}=\frac{\omega}{c}\mathbf{B},\qquad \mathbf{k}\times\mathbf{H}=-\frac{\omega}{c}\mathbf{D}
\label{eq:kcross}
\end{equation}
where $\mathbf{B}=\underline{\pmb{\mu}}\cdot\mathbf{H}$ and $\mathbf{D}=\underline{\pmb{\varepsilon}}\cdot\mathbf{E}$. \cref{eq:kcross} is a system of six linear equations with six unknowns, four of which are independent. This leads to the following 4$\times$4 matrix equation:

\begin{equation}\label{eq:3Dk}
q \begin{pmatrix}
											E_x\\
											E_y\\
											H_x\\
											H_y
											\end{pmatrix} =\begin{pmatrix}
										\underline{\pmb{0}}&\underline{\pmb{M}}_a\\
										\underline{\pmb{M}}_b&\underline{\pmb{0}}
																		\end{pmatrix}\begin{pmatrix}
											E_x\\
											E_y\\
											H_x\\
											H_y
											\end{pmatrix},
\end{equation}
where the non-zero quadrants are given by
\[\underline{\pmb{M}}_a=\]
\[
\begin{psmallmatrix}
			\tfrac{\omega}{c}\sin\theta\cos\theta(\mu_{1x}-\mu_{1y})&{-}\color{black}k_\rho^2\frac{c}{\varepsilon_{1y}\omega}{+}\color{black}\tfrac{\omega}{c}(\mu_{1y}\cos^2\theta+\mu_{1x}\sin^2\theta)\\
{-}\color{black}\tfrac{\omega}{c}(\mu_{1x}\cos^2\theta+\mu_{1y}\sin^2\theta)&\tfrac{\omega}{c}\sin\theta\cos\theta(\mu_{1y}-\mu_{1x})
			\end{psmallmatrix}
\]
and
\pagebreak
\[
\underline{\pmb{M}}_b=\]
\[\begin{psmallmatrix}
	\tfrac{\omega}{c}\sin\theta\cos\theta(\varepsilon_{1y}-\varepsilon_{1x})&{-}\color{black}\tfrac{\omega}{c}(\varepsilon_{1y}\cos^2\theta+\varepsilon_{1x}\sin^2\theta){+}\color{black}k_\rho^2\tfrac{c}{\mu_{1y}\omega}\\
	\tfrac{\omega}{c}(\varepsilon_{1x}\cos^2\theta+\varepsilon_{1y}\sin^2\theta)&\tfrac{\omega}{c}\sin\theta\cos\theta(\varepsilon_{1x}-\varepsilon_{1y}).
\end{psmallmatrix},
\]
{where $k_\rho$ denotes the radial component of the wavevector.}

The (extra)ordinary mode eigenvectors of \cref{eq:3Dk}, characterized by the  respective subscripts $e$ ad $o$, are 

\begin{subequations}
\begin{gather}
\pmb{\psi}_{o}=N_o^{-1}
\begin{psmallmatrix}
	{-}\color{black}\mu_{1y} q_{1o}\frac{\omega}{c}\sin\theta\\
	{\mu_{1y}q_{1o}\frac{\omega}{c}\cos\theta}\\
	k_{1z}^2\cos\theta\\
	-\varepsilon_{1y}\mu_{1y}\frac{\omega^2}{c^2}\sin\theta
\end{psmallmatrix},\\
\pmb{\psi}_{e}=N_e^{-1}
\begin{psmallmatrix}
	{-}\color{black}k_{1z}^2\cos\theta\\
	{-}\color{black}\varepsilon_{1y}\mu_{1y}\frac{\omega^2}{c^2}\sin\theta\\
	\varepsilon_{1y}q_{1e}\frac{\omega}{c}\sin\theta\\
	-\varepsilon_{1y}q_{1e}\frac{\omega}{c}\cos\theta
\end{psmallmatrix}
\end{gather}\label{eq:vec_eo}
\end{subequations}
where $N_{o,e}$ denote normalisation constants,  $q_{1o}=\sqrt{\varepsilon_{1y}\mu_{1x}\omega^2/c^2-(\mu_{1x}/\mu_{1y})k^2_\rho\cos^2\theta-k^2_\rho\sin^2\theta}$ and $q_{1e}=\sqrt{\varepsilon_{1x}\mu_{1y}\omega^2/c^2-(\varepsilon_{1x}/\varepsilon_{1y})k_\rho^2\cos^2\theta-k_\rho^2\sin^2\theta}$ are the respective eigenvalues,\footnote{The eigenvalues obtained in this work and in Ref. \cite{Lekner1991} do not agree with those obtained in Ref. \cite{RosaPRA2008}.} and we use the shorthand notation  $k_{1z}^2\equiv{\mu_{1y}}\color{black}{\varepsilon_{1y}\omega^2/c^2-k_\rho^2}$. \rem{As expected, the first two components of each eigenvector in \cref{eq:vec_eo} match Eq. (27) from Ref. \cite{Lekner1991} for $\mu_{1y}=1=\mu_{1x}$.} Note that only forward propagating modes are considered here, characterized by a positive real part, because only a single interface is taken into account for simplicity. The eigenvectors \cref{eq:vec_eo} determine the ratio between the electromagnetic field components inside the anisotropic medium. They will be required to obtain the Fresnel reflection coefficients. We {emphasize} that \cref{eq:vec_eo} should not be 
{conflated} with the $s$- and $p$-polarized modes{, which can be defined as the eigenmodes of \cref{eq:3Dk} for the isotropic case. These modes are characterized by the conditions $\{E_x\neq0, E_y=0, H_x=0, H_y\neq0\}$  and $\{E_x=0, E_y\neq0, H_x\neq0, H_y=0\}$, respectively. Clearly these conditions do not hold in this case. However, $s$- and $p$-mode decomposition to calculate the Casimir torque can still be found in the literature.} 

\section{Fresnel Reflection Coefficients} 

The anisotropic planar medium is faced with vacuum. As in any isotropic medium, in vacuum the electromagnetic waves can be written as a linear combination of $s$-polarized and $p$-polarized modes. This applies to both the incoming and the reflected waves. Inside the anisotropic medium, the fields can be written as a linear combination of the ordinary and extraordinary waves. The coefficients of these linear combinations are the Fresnel reflection and transmission coefficients. The condition of continuity of the tangential electromagnetic field components ensures that these coefficients are uniquely determined.

First, let the incoming wave be an $s$-wave, $\pmb{\psi}_{i,s}$. Then this wave will be reflected as a linear combination between $s$- and $p$-modes, denoted by $\pmb{\psi}_{rs}$. Hence the total wave in the isotropic media is $\pmb{\psi}_{i,s}+\pmb{\psi}_{rs}$. The first term is associated with an eigenvalue with a positive real part, whereas the eigenvalue of the latter mode has a negative real part.

Inside the anisotropic crystal, the electromagnetic wave can be written as a linear combination of ordinary and extraordinary waves. The ratio between the electromagnetic field components is fixed and the only degree of freedom is the proportionality constant for each mode. These constants are the transmission coefficients that couple the $s$-polarized wave to the ordinary and extraordinary waves: $\pmb{\psi}_{ts}=E_{y,i}(t_{so}\pmb{\psi}_{o}+t_{se}\pmb{\psi}_{e})$ where the eigenvectors $\pmb{\psi}_{o,e}$ are given by Eqs. \eqref{eq:vec_eo}, and the amplitude of the incoming wave $E_{y,i}>0$. Here the subscript $ts$ denotes the transmitted wave, that originates from an incoming $s$-wave, but is in itself \emph{not} an $s$-wave.  

The condition of continuity of the tangential electromagnetic field components at the interface implies that fields must be equal on both sides of the interface $\pmb{\psi}_{i,s}+\pmb{\psi}_{rs}=\pmb{\psi}_{ts}$, which leads to the a system of four equations with four unknowns. The relevant solutions are denoted by $r_{ss}=\tfrac{r_{1ssN}}{r_{1D}}$ and $r_{sp}=\frac{r_{1spN}}{r_{1D}}$ and they can be obtained in terms of the eigenvector components $\psi_{ok}$ and $\psi_{ek}$ ($k=1..4$). The procedure is now repeated for an incoming $p$-polarized wave. This leads to another system of four equations with four unknowns. The relevant solutions of this system are denoted by $r_{pp}=\tfrac{r_{1ppN}}{r_{1D}}$ and $r_{ps}=\tfrac{r_{1psN}}{r_{1D}}$. This leads to the following entries of Fresnel reflection matrix:

\begin{subequations}\label{eq:Fresnel0}
\begin{gather}
r_{1ssN}=k_0^2 (\psi_{e2} \psi_{o4}-\psi_{e4} \psi_{o2})+\\\nonumber
\qquad k_0 \tfrac{\omega}{c}  (\psi_{e1} \psi_{o2}-\psi_{e2} \psi_{o1}-\psi_{e3} \psi_{o4}+\psi_{e4} \psi_{o3})\\\nonumber
\qquad+\tfrac{\omega ^2}{c^2} (\psi_{e3} \psi_{o1}-\psi_{e1} \psi_{o3}),\\
r_{1spN}=2 k_0 \tfrac{\omega}{c}  (\psi_{e1} \psi_{o4}-\psi_{e4} \psi_{o1}),\\
r_{1ppN}=k_0^2 (\psi_{e4} \psi_{o2}-\psi_{e2} \psi_{o4})+\\\nonumber
\quad k_0 \tfrac{\omega}{c}  (\psi_{e1} \psi_{o2}-\psi_{e2} \psi_{o1}-\psi_{e3} \psi_{o4}+\psi_{e4} \psi_{o3})+\\\nonumber
\quad\tfrac{\omega ^2}{c^2} (\psi_{e1} \psi_{o3}-\psi_{e3} \psi_{o1}),\\
r_{1psN}=2  k_0 \tfrac{\omega}{c}  (\psi_{e2} \psi_{o3}-\psi_{e3} \psi_{o2}),\\
r_{1D}= k_0^2 (\psi_{e2} \psi_{o4}-\psi_{e4} \psi_{o2})\label{eq:rD}\\\nonumber
\quad+k_0 \tfrac{\omega}{c}  (\psi_{e1} \psi_{o2}-\psi_{e2} \psi_{o1}+\psi_{e3} \psi_{o4}-\psi_{e4} \psi_{o3})\\\nonumber
\quad+\tfrac{\omega ^2}{c^2} (\psi_{e1} \psi_{o3}-\psi_{e3} \psi_{o1}),
\label{eq:rpsN}
\end{gather}
\end{subequations}
where $k_0=\sqrt{\omega^2/c^2-k_\rho^2}$ denotes the $z$-component of the wavevector in vacuum. Note that the reflection coefficients in terms {of} the Maxwell eigenvector components, \cref{eq:Fresnel0} are not restricted to uniaxial materials, but  are valid for biaxial half spaces as well. Of course, the explicit form of the eigenvector components will change in the biaxial case. In particular we note that the reflection matrix is symmetric if and only if $\psi_{e1} \psi_{o4}-\psi_{e4} \psi_{o1}=\psi_{e2} \psi_{o3}-\psi_{e3} \psi_{o2}$, which holds in the uniaxial case. Also note that the normalization constants cancel out of the reflection matrix.

\rem{Inserting the components of the eigenvectors \cref{eq:vec_eo} into \cref{eq:Fresnel0} leads to explicit expressions of the reflection matrix elements in terms of the frequency and wavevector components. For the non-magnetic case, where $\mu_{1y}=1$ and $q_{1o}=k_{1z}$, these expressions match Eqs. (34) and (42) of Ref. \cite{Lekner1991}.}
{Inserting the eigenvectors \cref{eq:vec_eo} into \cref{eq:Fresnel0} yields the explicit reflection coefficients}

\begin{subequations}\label{eq:Fresnel1}
\begin{gather}
r_{1ssN}=\frac{\varepsilon_{1y}\mu_{1y}\omega ^2(k_0 \mu_{1y}-q_{1e}) (k_0 \varepsilon_{1y}+q_{1o})\sin ^2\theta}{c^2}\\\nonumber
\qquad- (k_0 q_{1e} \varepsilon_{1y}+k_{1z}^2)(k_{1z}^2-k_0 \mu_{1y}q_{1o})\cos ^2\theta\\
\label{eq:rss2}
r_{1spN}=\frac{k_0 \varepsilon_{1y} \mu_{1y} \omega (k_{1z}^2-q_{1e} q_{1o})\sin 2\theta}{c}=r_{1psN}\\
\label{eq:rsp2}
r_{1ppN}=(k_{1z}^2-k_0 q_{1e} \varepsilon_{1y})(k_0 \mu_{1y} q_{1o}+k_{1z}^2)\cos ^2\theta \\\nonumber
\qquad+\frac{\varepsilon_{1y}\mu_{1y} \omega ^2\sin ^2\theta  (k_0 \mu_{1y}+q_{1e}) (q_{1o}-k_0 \varepsilon_{1y})}{c^2}\\
\label{eq:rpp2}
r_{1D}=(k_0 q_{1e} \varepsilon_{1y}+k_{1z}^2)(k_0 \mu_{1y} q_{1o}+k_{1z}^2)\cos ^2\theta\\\nonumber
\qquad+\frac{ \varepsilon_{1y}\mu_{1y} \omega ^2 \sin ^2\theta  (k_0 \mu_{1y}+q_{1e}) (k_0 \varepsilon_{1y}+q_{1o})}{c^2}.
\label{eq:rD2}
\end{gather}
\end{subequations}
{For the non-magnetic case, where $\mu_{1y}=1$ and $q_{1o}=k_{1z}$, the results \cref{eq:Fresnel1} match Eqs. (34) and (42) of Ref. \cite{Lekner1991}.}

The equivalent reflection coefficients for a non-identical second medium can be easily obtained by transforming \cref{eq:Fresnel1} as follows: $q_{1e}\rightarrow{}q_{2e}$, $q_{1o}\rightarrow{}q_{2o}$, $k_{1z}\rightarrow k_{2z}$,$\varepsilon_{1y}\rightarrow\varepsilon_{2y}$, $\mu_{1y}\rightarrow\mu_{2y}$, and $\theta\rightarrow\theta+\varphi$, where $k_{2z}$, $q_{2 e,o}$, $\varepsilon_{2y}$,$\mu_{2y}$ represent the equivalents of $q_{1e,o}$, $k_{1z}$, $\varepsilon_{1y}$, $\mu_{1y}$, respectively for medium 2, and $\varphi$ represents the angle between the optic axes of the media. (See \cref{fig:fig1}). Next, the transformed eigenvectors must be inserted into \cref{eq:Fresnel0}.  In general one must take into account the different propagation directions for each medium, but in this case, but in this case the reflection matrix is invariant under $\mathbf{k}\rightarrow-\mathbf{k}$. 

\section{Casimir Energy and Torque} 

\noindent Now we are in a position to determine the Casimir torque. 
The Casimir energy per unit area is given by the Lifshitz formula~\cite{Lambrecht2006}

\begin{equation}
\begin{split}
\frac{E_{Cas}(a,\varphi)}{A}&=\frac{k_b T}{4\pi^2} \sum_{n=0}^{\infty}(1-\tfrac{1}{2}\delta_{n,0})\times\\
&\int_0^{\infty}\int_0^{2\pi}{\log(D(a,\mathbf{k},\theta,\varphi,i\zeta_n))}k_\rho\ud k_\rho\ud\theta
\end{split}
\label{eq:Lifshitz}
\end{equation}
with
\begin{equation}
\begin{split}
&D(a,\mathbf{k},\theta,\varphi,i\zeta_n)=\\
&\det\Big(\mathbf{\underline{I}}-\mathbf{\underline{r}_1}(\mathbf{k},\theta,i\zeta_n)\cdot\mathbf{\underline{r}_2}(\mathbf{k},\theta+\varphi,i\zeta_n)e^{-2k_0 a}\Big)
\end{split}
\label{eq:Dn}
\end{equation}
where $\mathbf{\underline{I}}$ denotes the 2$\times$2 identity matrix and $\mathbf{\underline{r}}_j$, $j=1,2$ represent the reflection matrices given by $\mathbf{\underline{r}}_j=\tfrac{1}{r_{jD}}
\begin{psmallmatrix}
	r_{j,ss,N}&r_{j,sp,N}\\
	r_{j,ps,N}&r_{j,pp,N}
\end{psmallmatrix},$
the elements of which are given by \cref{eq:Fresnel0} for $j=1$, and they should be transformed from medium 1 to 2 for $j=2$. For numerical convergence, all quantities are evaluated at the imaginary Matsubara frequencies $\zeta_n\equiv\frac{2\pi nk_bT}{\hbar}$ at finite temperature $T$. The Casimir torque is 

\begin{equation}
\tau(a,\varphi)=-\frac{\partial E_{Cas}}{\partial \varphi}.
\label{eq:torque}
\end{equation}

{The form of \cref{eq:Dn} brings us to the apparent symbolic difference between this result and that of Ref. \cite{Philbin2008}. First note that $D$ is a quadratic function of $\exp(-2k_0a)$ and  let us introduce the notation}

\begin{equation}
\begin{split}
&D(\mathbf{k},\varphi,a)\equiv\\
&1+P(\mathbf{k},\varphi)\exp(-2k_0a)+Q(\mathbf{k},\varphi)\exp(-4k_0a).
\end{split}
\label{eq:PQNot}
\end{equation}
{If it is assumed that the reflection matrices are symmetric, (which they are in the birefringent case), the numerator of $P$ can be written as}

\begin{equation}
\begin{split}
&r_{1D}r_{2D}P=\\
&-r_{1ppN} r_{2ppN}-2 r_{1spN} r_{2spN}-r_{1ssN} r_{2ssN},
\end{split}
\label{eq:55}
\end{equation}
{and that of $Q$ is}
\begin{equation}
\begin{split}
&r_{1D}^2r_{2D}^2Q=\\
&(r_{1spN}^2-r_{1ppN} r_{1ssN}) (r_{2spN}^2-r_{2ppN} r_{2ssN}).
\end{split}
\label{eq:56}
\end{equation}
{ The forms of \cref{eq:55,eq:56} are identical to those of Eqs. (55) and (56), respectively of Ref. \cite{Philbin2008}. So at least in terms of the entries of the reflection matrices, there does not appear to be a symbolic difference between the result of this work and that of Ref.~\cite{Philbin2008}.}

Next we will {compare  this result for $\mu_{1x}=1=\mu_{1y}$ to that of Ref. \cite{Barash1978}}.

\section{Summary of proof of Barash's formula}

\begin{figure}[!htbp]
	\centering
				\includegraphics[width=0.48\textwidth]{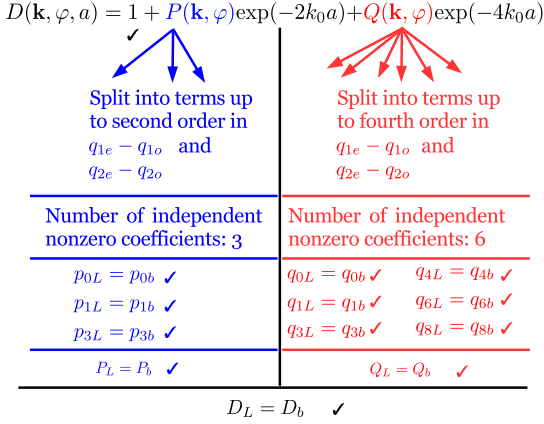}
	\caption{Overview of the proof of Barash's fomula.  The coefficients $P_{b,L}$ and $Q_{b,L}$ can be written as  $P_{b,L}=p_{0b,L}+p_{1,b,L}(q_{1e}-q_{1o})+p_{2b,L} (q_{2e}-q_{2o})+p_{3b,L}(q_{1e}-q_{1o})(q_{2e}-q_{2o})$ and $Q_{b,L}=q_{0b,L}+q_{1b,L}(q_{1e}-q_{1o})+q_{2b,L}(q_{2e}-q_{2o})+q_{3b,L}(q_{1e}-q_{1o})(q_{2e}-q_{2o})+q_{4b,L}(q_{1e}-q_{1o})^2+q_{5b,L}(q_{2e}-q_{2o})^2+q_{6b,L}(q_{1e}-q_{1o})(q_{2e}-q_{2o})^2+q_{7b,L} (q_{1e}-q_{1o})^2(q_{2e}-q_{2o})+q_{8b,L}(q_{1e}-q_{1o})^2(q_{2e}-q_{2o})^2$. The subscripts $b$ and $L$ denote results from Ref.~\cite{Barash1978} and from this work, respectively. Symmetry between the labels of the media reduces the number of independent coefficients.}
	\label{fig:fig2}
\end{figure}

{The question that arises now is: how does the result \cref{eq:torque} for non-magnetic materials compare to Eq. (27) of Barash's paper \cite{Barash1978}? A detailed comparison can be found in the appendix. We will provide a summary here, omitting the algebraic details. We contend that \cref{eq:torque} for non-magnetic materials is identical to Eq. (27) of Ref.~\cite{Barash1978}. This was {claimed} in Ref.~\cite{Somers2017} without proof. The basic idea of the proof is that this large, complicated problem is split into smaller, simpler and independent parts. This is illustrated by \cref{fig:fig2}. }  

{To address this problem, the first step is the realization that this comparison boils down to that between the arguments of the logarithms, i.e. $D$ of \cref{eq:Dn} to Eq. (21) of Ref. \cite{Barash1978}. Eq. (1) of Ref. \cite{Barash1978}, which defines the Casimir energy as the Helmholtz free energy, is identical to the Matsubara sum in the Lifshitz formula \cref{eq:Lifshitz}. For the sake of this comparison we introduce the subscripts $b$ and $L$, denoting results from Ref. \cite{Barash1978} and this work respectively. Hence the comparison can be limited to that between $D_b$ and $D_L$.} 

{The next step is the observation that both $D_b$ and $D_L$ are quadratic functions of $\exp(-2k_0a)$. (Cf. \cref{eq:PQNot}.) This reduces the comparison to that between the coefficients of $D_b$ and $D_L$. }

{This brings us to the constant (distance independent) term. This corresponds to the limit of large distances where $k_0a\rightarrow\infty$. It is expected that both $D_b$ and $D_L$ tend to unity as $k_0a\rightarrow\infty$, because the Casimir energy must tend to zero in this limit. From \cref{eq:Dn} it can be easily seen that $D_L\rightarrow1$, but that $D_b\rightarrow1$ is less obvious. The proof of the latter can be found in \cref{Barash1} of the Appendix.}

{Here we continue to use the notation of \cref{eq:PQNot}, but now with the subscripts $b$ and $L$. It is important to realize that Barash wrote the coefficients in terms of the differences between the ordinary and extraordinary mode eigenvalues. More specifically, $P_b$ is a second degree polynomial in $q_{ie}-q_{io}$ with $i=1,2$ and $Q_b$ is a fourth degree polynomial in the same variables. Hence in order to come to a comparison, $P_L$ and $Q_L$ from \cref{eq:PQNot} must be written in the same way. Let us introduce the following notation for the coefficients for both Barash's and Lifshitz's versions of $P$ and $Q$:  $P_{b,L}=p_{0b,L}+p_{1,b,L}(q_{1e}-q_{1o})+p_{2b,L} (q_{2e}-q_{2o})+p_{3b,L}(q_{1e}-q_{1o})(q_{2e}-q_{2o})$ and $Q_{b,L}=q_{0b,L}+q_{1b,L}(q_{1e}-q_{1o})+q_{2b,L}(q_{2e}-q_{2o})+q_{3b,L}(q_{1e}-q_{1o})(q_{2e}-q_{2o})+q_{4b,L}(q_{1e}-q_{1o})^2+q_{5b,L}(q_{2e}-q_{2o})^2+q_{6b,L}(q_{1e}-q_{1o})(q_{2e}-q_{2o})^2+q_{7b,L} (q_{1e}-q_{1o})^2(q_{2e}-q_{2o})+q_{8b,L}(q_{1e}-q_{1o})^2(q_{2e}-q_{2o})^2$, where the comma denotes that one can choose between the subscripts $b$ or $L$. }

{Before moving on to the direct comparison between coefficients of $P_L$,  $Q_L$ and $P_b$ and $Q_b$ respectively, it is worth noting that the labels 1 and 2 of the media and their respective angles $\theta$ and $\theta+\varphi$ are arbitrary and interchanging them should not change the physics. This symmetry leads to the useful relations \cref{eq:p1p2,eq:q1q2,eq:q4q5,eq:q6q7}, which reduce the number of independent coefficients from 4 to 3 for $P_{b,L}$ and from 9 to 6 for $Q_{b,L}$. (See \cref{fig:fig2}.)}

{Another step is now to rid Barash's expression of the fractions within its numerator and denominator. This is done by multiplying numerator and denominator of both $P_b$ and $Q_b$ by the factor $(q_{1o}^2+k_\rho^2\sin^2\theta)(q_{2o}^2+k_\rho^2\sin^2(\theta+\varphi))$. Now, the denominators of $P_b$ and $P_L$ can be related as follows }

\begin{equation}
r_{1D}r_{2D}=(q_{1o}^2+k_\rho^2\sin^2\theta)(q_{2o}^2+k_\rho^2\sin^2(\theta+\varphi))\gamma 
\label{eq:denominators}
\end{equation}
{where $\gamma$, the denominator of $P_b$, is given by \cref{eq:gamma}. Here we have taken advantage of the ordinary mode eigenvalue relations}

\begin{equation}
q_{jo}^2+k_\rho^2=\varepsilon_{jy} (k_0^2+k_\rho^2)\qquad j=1,2.
\label{eq:qjo}
\end{equation}
{Note that \cref{eq:qjo} is simply the combination of the definitions of $q_{io}$ and $k_0$ with $\omega/c$ eliminated. With this elimination we follow Ref. \cite{Barash1978}.  }

{Since \cref{eq:denominators} tells us how to relate the denominators of $P_b$ ad $P_L$, we are now ready to compare their numerators.  The comparison is greatly simplified by assuming $q_{1e}=q_{1o}$ and $q_{2e}=q_{2o}$. This immediately establishes that $p_{0b}=p_{0L}$. We proceed with the assumption $q_{1e}\neq q_{1o}$ and $q_{2e}=q_{2o}$. The already established coefficient $p_{0L}$ is subtracted from the resulting expression for $P_L$ under this condition. This will yield and expression for $p_{1L}$, which is indeed identical to $p_{1b}$ if \cref{eq:qjo} is taken into account. The symmetry relation \cref{eq:p1p2} allows us to skip $p_{2L}$, and we move on to $p_{3L}$. This requires the general case  $q_{1e}\neq q_{1o}$ and $q_{2e}\neq q_{2o}$. From the full expression of $P_L$ the other terms that have been obtained so far, are subtracted, i.e. $P_L-(p_{0L}+p_{1L}(q_{1e}-q_{1o})+p_{2L}(q_{2e}-q_{2o}))$. The expression for $p_{3L}$ found in this way is equal to $p_{3b}$ with the eigenvalue relations \cref{eq:qjo}. The detailed proof that $P_b=P_L$ can be found in \cref{sec:coeffP}.}

{Now we can proceed with the coefficient $Q$. The denominator of $Q_L$ is the square of that of $P_L$. The numerator of  $Q_b$ is multiplied with $(q_{1o}^2+k_\rho^2\sin^2\theta)(q_{2o}^2+k_\rho^2\sin^2(\theta+\varphi))\gamma $, so that both $Q_b$ and $Q_L$ have the same denominator. Consequently, $Q_b$ and $Q_L$ now have the number of coefficients. Most of these coefficients can be compared in the same way as those of $P_{b,L}$. However, for the mixed terms, the trick with the simplifying assumptions $q_{1e}=q_{1o}$ or $q_{2e}=q_{2o}$ no longer works. In this case the expressions for the coefficients $Q_L$ must be fully expanded in order to write them in the desired form. The full proof that $Q_L=Q_b$ can be found in \cref{sec:CoeffQ}.}

\rem{\subsection{Proof of Barash's formula} 

We contend that \cref{eq:torque} for non-magnetic materials is identical to Eq. (27) of Ref.~\cite{Barash1978}. This was 
{claimed} in Ref.~\cite{Somers2017} without proof. For a detailed proof of this statement, we refer to the Appendix. 
Here we will give a brief summary. Essentially, the proof can be simplified by splitting it up into simpler, smaller independent parts. That is, the argument of the logarithm in the Lifshitz formula $D$ is a polynomial in $\exp(-2k_0a)$: $D(\mathbf{k},\varphi,a)=1+P_n(\mathbf{k},\varphi)\exp(-2k_0a)+Q_n(\mathbf{k},\varphi)\exp(-4k_0a)$. The coefficients $P$ and $Q$  are in turn  multivariate polynomials in $q_{1e}-q_{1o}$ and $q_{2e}-q_{2o}$. (See \cref{fig:fig2} for a schematic overview). Most of the coefficients of these multivariate polynomials can be determined with the simplifying assumptions $q_{1e}=q_{1o}$ and $q_{2e}=q_{2o}$. In this way, the coefficients according to Barash and \cref{eq:Dn} can be compared. It turns out that they do indeed match, { if the eigenvalue conditions $q_{jo}^2=\varepsilon_{jy}(k_0^2-k_\rho^2)$, $j=1,2$ are taken into account.}  To our knowledge, this is the first independent {analytic verification} 
of the result of Ref.~\cite{Barash1978}.}

\section{Conclusions} We have derived an exact {and general} expression for the Casimir torque between two half-spaces that exhibit both electric and magnetic birefringence, {and thereby analytically verified the Casimir torque result of Ref.~\cite{Barash1978} as a limiting case of non-magnetically permeable media. 
This is the first complete and transparent analytic verification in the more than forty intervening years since the publication of the results of Ref.~\cite{Barash1978}. 

Our approach {does not depend on approximations such as weak anisotropy or neglecting retardation effects, and} 
can be generalized to other types of media, such as metamaterials~\cite{Poddubny2013}, multiferroics~\cite{Fiebig2016},  topological matter~\cite{Tokura2019}, biaxial materials~\cite{Thiyam2018}, and non-reciprocal materials~\cite{Fuchs2017}. The recently reported measurements of the Casimir torque~\cite{Somers2018} should instigate further investigations for which this result is of {considerable} importance. 

\begin{acknowledgments}

WB thanks B. J. Hoenders for a useful email exchange. We acknowledge support from a start-up grant from Nanyang Technological University, under the number M4082095.110.
\end{acknowledgments}

\bibliography{Casimir3}

\newpage

\widetext

\appendix 

\section{Comparison to Barash's result}

\subsection{Introduction}
\begin{table}[!hbtp]
\begin{tabular}{|l|l|}
\hline
Ref. \cite{Barash1978}& This document\\ \hline
$u=r\cos\varphi$ &$k_x=k_\rho\cos\theta$  \\ \hline
$v=r\sin\varphi$ &$k_y=k_\rho\sin\theta$  \\ \hline
$\varphi$& $\theta$ \\ \hline
$r$&$k_\rho$  \\ \hline
$\theta$ & $\varphi$ \\ \hline
$\rho_1$ &$iq_{1o}$  \\ \hline
$\rho_2$ &$iq_{2o}$  \\ \hline
$\rho_1'(\theta)$ &$iq_{1e}(\varphi)$  \\ \hline
$\rho_2'(\theta)$&$iq_{2e}(\varphi)$ \\ \hline
$\rho_3$ &$ik_0$\\ \hline
$\varepsilon_{\mathrm{I } 1}$ &$\varepsilon_{1x}$ \\\hline
$\varepsilon_{\mathrm{I } 2}$ &$\varepsilon_{1y}$ \\\hline
$\varepsilon_{\mathrm{II } 1}$ &$\varepsilon_{2x}$ \\\hline
$\varepsilon_{\mathrm{II } 2}$ &$\varepsilon_{2y}$ \\\hline
$\varepsilon_3$&1\\\hline
$l$&$a$\\\hline
\end{tabular}
\caption{ Comparison between the conventions of this document and those of Ref. \cite{Barash1978}. Unlike Ref. \cite{Barash1978} we do not include the possibility of a liquid gap. On the other hand, we allow the option of magnetic anisotropy. Hence it is assumed here that $\mu_{jx}=1=\mu_{jy}$. The $z$-components of the wave vector are purely imaginary since they are evaluated at imaginary frequencies. }
\label{tab:conventions}
\end{table}
\noindent 
{In this Appendix, we }
prove that the result obtained by Barash, Eq (21) in \cite{Barash1978}, is identical to the combination of the Lifshitz formula \cite{Lambrecht2006} with Lekner's reflection coefficients \cite{Lekner1991}, i.e. {\cref{eq:Dn}} for the case  $\mu_x=1=\mu_y$. To this end, it suffices to compare the argument of the logarithm in {\cref{eq:Dn}} , which Barash calls `dispersion equation', denoted here by $D$. \cite{Barash1978} After all, both results are written in the same form and the only thing that could be different is $D$.

Since the expression for $D$ {can be quite} 
complicated in both cases, it helps to split them into smaller, simpler parts. Firstly, note that it is a quadratic function of $\exp(-2k_0a)$: $D(\mathbf{k},\varphi,a)=1+P(\mathbf{k},\varphi)\exp(-2k_0a)+Q(\mathbf{k},\varphi)\exp(-4k_0a)$. (Part of this {Appendix}
will be dedicated to showing that the constant term of Barash's result equals unity indeed.) Next, the proof takes advantage of the fact that both $P$ and $Q$ are both polynomials in the two variables $q_{1e}-q_{1o}$ and $q_{2e}-q_{2o}$. The coefficients $P$ and $Q$ will be compared, by comparing the coefficients of these polynomials.

The notation of Ref.\cite{Barash1978} is {somewhat}
unconventional. For reference we have included a table (\cref{tab:conventions}) to clarify how it compares to our notation.

The {Appendix}
is organized as follows. After the introduction, both results will simply be given. Then it will be shown that Barash's version of $D$ tends to one at sufficiently large distances. Next, the symmetry of between the coefficients will be discussed, thereby reducing the number of independent coefficients. Finally, the coefficients of both $P$ and $Q$ will be compared.

\subsection{Barash vs. Lifshitz-Lekner's result}
\noindent 
In the notation of this document, Barash's result is

\begin{equation}
D_b=\frac{A_1-\frac{\varepsilon_{2y} (i q_{2e}-i q_{2o}) (B k_\rho^2 \sin ^2(\theta +\varphi )+C-E (2 k_\rho^2 \sin (\theta ) \cos (\varphi ) \sin (\theta +\varphi )-k_0^2 \sin ^2(\varphi )))}{-k_\rho^2 \sin ^2(\theta +\varphi )-q_{2o}^2}}{\gamma }
\label{eq:DB}
\end{equation}
where the subscript $b$ denotes that this is a result by Barash. The coefficients are given by

\[A_1=((i k_0+i q_{2o}) (i k_0+i q_{1o})-e^{-2 a k_0} (i q_{2o}-i k_0) (i q_{1o}-i k_0))\times\]
\[ ((i k_0 \varepsilon_{2y}+i q_{2o}) (i k_0 \varepsilon_{1y}+i q_{1o})-e^{-2 a k_0} (i q_{2o}-i k_0 \varepsilon_{2y}) (i q_{1o}-i k_0 \varepsilon_{1y}))\]
\[-\frac{1}{-k_\rho^2 \sin ^2(\theta )-q_{1o}^2}\Big(\varepsilon_{1y} (i q_{1e}-i q_{1o}) (2 (\varepsilon_{2y}-1) e^{-2 a k_0} (k_\rho^2 \sin ^2(\theta ) (i k_0^2 q_{2o}+i k_\rho^2 q_{1o})\]
\[-i k_0^2 q_{1o} (-2 k_\rho^2 \sin ^2(\theta )+k_\rho^2-q_{2o} q_{1o}))+e^{-4 a k_0} (i q_{2o}-i k_0) (i q_{1o}-i k_0) (i q_{2o}-i k_0 \varepsilon_{2y})\times\]
\[ (k_\rho^2 \sin ^2(\theta )-k_0 q_{1o})+(i k_0+i q_{2o}) (i k_0+i q_{1o}) (i k_0 \varepsilon_{2y}+i q_{2o}) (k_0 q_{1o}+k_\rho^2 \sin ^2(\theta )))\Big),\]

\[B=\frac{-1}{k_\rho^2 \sin ^2(\theta )+q_{1o}^2}\Big[\varepsilon_{1y} (i q_{1e}-i q_{1o}) (2 e^{-2 a k_0} (k_\rho^2 \sin ^2(\theta ) (-k_0^2-q_{2o} q_{1o})-k_0^2 q_{2o} q_{1o})\]
\[-e^{-4 a k_0} (i q_{2o}-i k_0) (i q_{1o}-i k_0) (k_\rho^2 \sin ^2(\theta )-k_0 q_{1o})\]
\[+(i k_0+i q_{2o}) (i k_0+i q_{1o}) (-k_0 q_{1o}-k_\rho^2 \sin ^2(\theta )))\Big]+2 (\varepsilon_{1y}-1) e^{-2 a k_0} (2 i k_0^2 q_{2o}+i k_0^2 q_{1o}+i k_\rho^2 q_{2o})\]
\[+e^{-4 a k_0} (i q_{2o}-i k_0) (i q_{1o}-i k_0) (i q_{1o}-i k_0 \varepsilon_{1y})+(i k_0+i q_{2o}) (i k_0+i q_{1o}) (i q_{2o} \varepsilon_{1y}+i q_{1o}),\]

\[C=\frac{1}{k_\rho^2 \sin ^2(\theta )+q_{1o}^2}\Big[k_0 q_{2o} \varepsilon_{1y} (i q_{1e}-i q_{1o}) (2 i k_0 e^{-2 a k_0} (-i k_0^2 q_{1o}+k_\rho^2 \sin ^2(\theta ) (i q_{1o}-i q_{2o})-i q_{2o} q_{1o}^2)\]
\[-e^{-4 a k_0} (i q_{2o}-i k_0) (i q_{1o}-i k_0) (k_\rho^2 \sin ^2(\theta )-k_0 q_{1o})+(i k_0+i q_{2o}) (i k_0+i q_{1o}) (k_0 q_{1o}+k_\rho^2 \sin ^2(\theta )))\Big]\]
\[-k_0 q_{2o} (2 i k_0 (\varepsilon_{1y}-1) e^{-2 a k_0} (k_\rho^2+q_{2o} q_{1o})+e^{-4 a k_0} (i q_{2o}-i k_0) (i q_{1o}-i k_0) (i q_{1o}-i k_0 \varepsilon_{1y})\]
\[+(i k_0+i q_{2o}) (i k_0+i q_{1o}) (-i k_0 \varepsilon_{1y}-i q_{1o})),\]

\[E=\frac{4 k_0^2 q_{2o} q_{1o} \varepsilon_{1y} e^{-2 a k_0} (i q_{1e}-i q_{1o})}{-k_\rho^2 \sin ^2(\theta )-q_{1o}^2},\]
and

\begin{equation}
\gamma=(i k_0+i q_{2o}) (i k_0+i q_{1o}) (-\frac{\varepsilon_{2y} (i q_{1e}-i q_{1o}) (k_0 q_{1o}+k_\rho^2 \sin ^2(\theta ))}{-k_\rho^2 \sin ^2(\theta )-q_{1o}^2}+i k_0 \varepsilon_{1y}+i q_{1o})\times
\label{eq:gamma}
\end{equation}
\[  (-\frac{\varepsilon_{2y} (i q_{2e}-i q_{2o}) (k_0 q_{2o}+(k_\rho \sin (\theta ) \cos (\varphi )+k_\rho \cos (\theta ) \sin (\varphi ))^2)}{-(k_\rho \sin (\theta ) \cos (\varphi )+k_\rho \cos (\theta ) \sin (\varphi ))^2-q_{2o}^2}+i k_0 \varepsilon_{2y}+i q_{2o})
\]

 The argument of the logarithm in Lifshitz' formula {\cref{eq:Dn}}  in terms of the entries of the reflection matrix  is

\[D_L=\frac{ e^{-4 a k_0}}{r_{D1}^2 r_{D2}^2}\Big(r_{ppN1} r_{ppN2} r_{ssN1} r_{ssN2}-r_{ppN1} r_{psN2} r_{spN2} r_{ssN1}\]
\[-r_{ppN2} r_{psN1} r_{spN1} r_{ssN2} +r_{psN1} r_{psN2} r_{spN1} r_{spN2}\Big)\]
\[-\frac{e^{-2 a k_0}}{r_{D1} r_{D2}}\Big(r_{ppN1} r_{ppN2} +r_{psN1} r_{spN2}+r_{psN2} r_{spN1} +r_{ssN1} r_{ssN2} \Big)+1,\]
where the subscript $L$ denotes that this is a combination of results by Lifshitz \cite{Lifshitz55,Lifshitz61} and Lekner. \cite{Lekner1991} 
%

The entries of the reflection matrix from Eq. (7) from the main paper  for $\mu_y=1$ and $k_z=q_{1o}$  are rewritten  as follows: 

\[r_{ssN1}=\sin ^2(\theta ) ((k_0-q_{1e}) (k_\rho^2+q_{1o}^2) (k_0 \varepsilon_{1y}+q_{1o})-q_{1o} (k_0-q_{1o}) (k_0 q_{1e} \varepsilon_{1y}+q_{1o}^2))+q_{1o} (k_0-q_{1o}) (k_0 q_{1e} \varepsilon_{1y}+q_{1o}^2)\]

\[r_{ppN1}=\sin ^2(\theta ) (-(k_0+q_{1e}) (k_\rho^2+q_{1o}^2) (k_0 \varepsilon_{1y}-q_{1o})-q_{1o} (k_0+q_{1o}) (q_{1o}^2-k_0 q_{1e} \varepsilon_{1y}))+q_{1o} (k_0+q_{1o}) (q_{1o}^2-k_0 q_{1e} \varepsilon_{1y})\]

\[r_{spN1}=k_0 q_{1o} \varepsilon_{1y} \sin (2 \theta ) \sqrt{k_0^2+k_\rho^2} (q_{1o}-q_{1e})=r_{psN1}\]

\[r_{D1}=\sin ^2(\theta ) ((k_0+q_{1e}) (k_\rho^2+q_{1o}^2) (k_0 \varepsilon_{1y}+q_{1o})-q_{1o} (k_0+q_{1o}) (k_0 q_{1e} \varepsilon_{1y}+q_{1o}^2))+q_{1o} (k_0+q_{1o}) (k_0 q_{1e} \varepsilon_{1y}+q_{1o}^2),\]
where we have used the relations $k_0=\sqrt{\omega^2/c^2-k_\rho^2}$ and $q_{1o}=\sqrt{\varepsilon_{1y}\omega^2/c^2-k_\rho^2}$, the latter of which holds only in the non-magnetic case. To obtain the analogous expressions for the second medium, the substitutions from from one medium to the other  must be performed.

\subsection{Barash's result in the limit of large distances}\label{Barash1}

\noindent At sufficiently large distances, the exponentially decreasing terms vanish and only the constant term remains. The coefficients $A$ to $C$ in \cref{eq:DB} are in this limit

\[A_{\lim}=(i k_0+i q_{2o}) (i k_0+i q_{1o}) (i k_0 \varepsilon_{2y}+i q_{2o}) (i k_0 \varepsilon_{1y}+i q_{1o})\]
\[-\frac{\varepsilon_{1y} (i k_0+i q_{2o}) (i k_0+i q_{1o}) (i q_{1e}-i q_{1o}) (i k_0 \varepsilon_{2y}+i q_{2o}) (k_0 q_{1o}+k_\rho^2 \sin ^2(\theta ))}{-k_\rho^2 \sin ^2(\theta )-q_{1o}^2}\]

\[B_{\lim}=\frac{\varepsilon_{1y} (i k_0+i q_{2o}) (i k_0+i q_{1o}) (i q_{1e}-i q_{1o}) (-k_0 q_{1o}-k_\rho^2 \sin ^2(\theta ))}{-k_\rho^2 \sin ^2(\theta )-q_{1o}^2}+(i k_0+i q_{2o}) (i k_0+i q_{1o}) (i k_0 \varepsilon_{1y}+i q_{1o})\]

\[C_{\lim}=-\frac{k_0 q_{2o} \varepsilon_{1y} (i k_0+i q_{2o}) (i k_0+i q_{1o}) (i q_{1e}-i q_{1o}) (k_0 q_{1o}+k_\rho^2 \sin ^2(\theta ))}{-k_\rho^2 \sin ^2(\theta )-q_{1o}^2}\]
\[-k_0 q_{2o} (i k_0+i q_{2o}) (i k_0+i q_{1o}) (-i k_0 \varepsilon_{1y}-i q_{1o}),\]
while $E_{\lim}=0$.

Now the numerator of $D_b$ for $k_0a\gg1$ is:

\[A_{\lim} -\frac{\varepsilon_{2y} (q_{2e}-q_{2o}) (B_{\lim} k_\rho^2 \sin ^2(\theta +\varphi )+C_{\lim})}{q_{2o}^2-k_\rho^2 \sin ^2(\theta +\varphi )}=\]
\[\frac{(k_0+q_{2o}) (k_0+q_{1o})}{(k_\rho^2 \sin ^2(\theta )+q_{1o}^2) (k_\rho^2 \sin ^2(\theta +\varphi )+q_{2o}^2)}\Big[ (-k_\rho^2 \sin ^2(\theta ) (q_{1o} (\varepsilon_{1y}-1)-\varepsilon_{1y} (k_0+q_{1e}))+k_0 q_{1e} q_{1o} \varepsilon_{1y}+q_{1o}^3)\times\]
\[ (-k_\rho^2 \sin ^2(\theta +\varphi ) (q_{2o} (\varepsilon_{2y}-1)-\varepsilon_{2y} (k_0+q_{2e}))+k_0 q_{2e} q_{2o} \varepsilon_{2y}+q_{2o}^3)\Big]\]
To obtain $D_b$ in this limit, this quantity must be divided by $\gamma$ from \cref{eq:gamma}, which does not depend on $a$. It can be seen that this numerator is actually the same as $\gamma$, hence

\[\lim_{a\rightarrow\infty}D_b=1.\]
Since $D_b$ is the argument of the logarithm in the Lifshitz formula {\cref{eq:Dn}}  the Casimir energy tends to zero as $a$ goes to infinity.

\subsection{Symmetry relations between coefficients}

\noindent We take advantage of the fact that $P_b$ is a second degree polynomial in $q_{1e}-q_{1o}$ and $q_{2e}-q_{2o}$, and that $Q_b$ is a fourth degree polynomial in the same variables. Generally a polynomial of degree $d$ with $n$ variables has $\binom{n+d}{d}=\binom{n+d}{n}$ coefficients. Hence in our case $P_b$ has 6 coefficients and $Q_b$ has 15. However, the number of nonzero coefficients are 4 and 9 respectively. Moreover, these coefficients are not independent of each other. Note that the labels 1 and 2 of the media are arbitrary, so switching them should not affect the torque. The constant term does not contribute to the torque and it can be ignored. Since medium 1 is associated with an optic axis with angle $\theta$ and medium 2 is associated with $\theta+\varphi$, these angles must be interchanged as well. Hence we define the following transformation

\begin{equation}
T: \{P(\theta,\theta+\varphi),Q(\theta,\theta+\varphi)\}\rightarrow
\label{eq:T}
\end{equation}
\[\{P(1\leftrightarrow2,\theta\leftrightarrow\theta+\varphi),Q(1\leftrightarrow2,\theta\leftrightarrow\theta+\varphi)\}\]
where `$1\leftrightarrow2$' denotes that the subscript 1 needs to be replaced by 2 and vice versa. In other words, the transformations of from one medium to the other  and their reverse have to be performed simultaneously. The condition that the torque must not change leads to 

\[P(1\leftrightarrow2,\theta\leftrightarrow\theta+\varphi)=P(\theta,\theta+\varphi)\]
\[Q(1\leftrightarrow2,\theta\leftrightarrow\theta+\varphi)=Q(\theta,\theta+\varphi)\]
Now let
\[P(\theta,\theta+\varphi)=p_0(\theta,\theta+\varphi)+p_1(\theta,\theta+\varphi)(q_{1e}-q_{1o})+p_2(\theta,\theta+\varphi)(q_{2e}-q_{2o})+p_3(\theta,\theta+\varphi)(q_{1e}-q_{1o})(q_{2e}-q_{2o})\]
Then we come to the following symmetry relations for the coefficients of $P$:

\[p_0(\theta,\theta+\varphi)=p_0(\theta+\varphi,\theta)\]
\begin{equation}
p_1(\theta,\theta+\varphi)=p_2(\theta+\varphi,\theta)
\label{eq:p1p2}
\end{equation}
\[p_3(\theta,\theta+\varphi)=p_3(\theta+\varphi,\theta)\]
And similarly for $Q$:

\[Q(\theta,\theta+\varphi)=q_0(\theta,\theta+\varphi)+q_1(\theta,\theta+\varphi)(q_{1e}-q_{1o})+\]
\[q_2(\theta,\theta+\varphi)(q_{2e}-q_{2o})+q_3(\theta,\theta+\varphi)(q_{1e}-q_{1o})(q_{2e}-q_{2o})+\]
\[q_4(\theta,\theta+\varphi)(q_{1e}-q_{1o})^2+ q_5(\theta,\theta+\varphi)(q_{2e}-q_{2o})^2+q_6(\theta,\theta+\varphi)(q_{1e}-q_{1o})^2(q_{2e}-q_{2o})\]
\[+q_7(\theta,\theta+\varphi)(q_{1e}-q_{1o})(q_{2e}-q_{2o})^2+q_8(\theta,\theta+\varphi)(q_{1e}-q_{1o})^2(q_{2e}-q_{2o})^2\]
we have
\[q_0(\theta,\theta+\varphi)=q_0(\theta+\varphi,\theta)\]
\begin{equation}
q_1(\theta,\theta+\varphi)=q_2(\theta+\varphi,\theta)
\label{eq:q1q2}
\end{equation}
\[q_3(\theta,\theta+\varphi)=q_3(\theta+\varphi,\theta)\]
\begin{equation}
q_4(\theta,\theta+\varphi)=q_5(\theta+\varphi,\theta)
\label{eq:q4q5}
\end{equation}
\begin{equation}
q_6(\theta,\theta+\varphi)=q_7(\theta+\varphi,\theta)
\label{eq:q6q7}
\end{equation}
\[q_8(\theta,\theta+\varphi)=q_8(\theta+\varphi,\theta).\]
The symmetry relations should be valid for both the Barash and the Lifshitz-Lekner versions of these coefficients. In the latter case, this is immediately seen from \cref{eq:PL,eq:QL}. 

In the former case it is not so obvious that this symmetry holds. Therefore this needs to be checked. Switching the labels on Barash's version of $D$ yields:

\[\frac{{A'}-\frac{{\varepsilon_{1y}} (i {q_{1e}}-i {q_{1o}}) ({B'} {k_\rho}^2 \sin ^2(\theta )+{C'}-{E'} (2 {k_\rho}^2 \sin (\theta ) \cos (\varphi ) \sin (\theta +\varphi )-{k_0}^2 \sin ^2(\varphi )))}{-{k_\rho}^2 \sin ^2(\theta )-{q_{1o}}^2}}{\gamma' }\]
where the primed quantities denote the transformed version of the unprimed quantities. In order to have $D'=D$ we require $\gamma'=\gamma$, which is relatively easily seen to hold.

Of course we are especially interested in symmetry relations between coefficients with different subscripts, because these could simplify some of the calculations. In particular we note that they reduce the number of independent coefficients from 4 to 3 for $P$ and from 9 to 6 for $Q$.

\subsection{The coefficient $P$}
\label{sec:coeffP}

\noindent Next, let us focus on the part of $D_L$ that is proportional to $\exp(-2k_0a)$, denoted by $P_L$:
\begin{equation}
P_L\equiv\frac{-1}{r_{D1} r_{D2}}\Big(r_{ppN1} r_{ppN2} +r_{psN1} r_{spN2}+r_{psN2} r_{spN1} +r_{ssN1} r_{ssN2} \Big)
\label{eq:PL}
\end{equation}
where $r_{psN1} r_{spN2}+r_{psN2} r_{spN1}$ can be simplified to $2r_{spN1}r_{spN2}$. Barash's equivalent of this coefficient is
\[P_b=\frac{P_A}{\gamma}+\]
\begin{equation}
\frac{i\varepsilon_{2y} (q_{2e}-q_{2o}) (-P_E (2 k_\rho^2 \sin (\theta ) \cos (\varphi ) \sin (\theta +\varphi )-k_0^2 \sin ^2(\varphi ))+k_\rho^2 P_B \sin ^2(\theta +\varphi )+P_C)}{\gamma(k_\rho^2 \sin ^2(\theta +\varphi )+q_{2o}^2)}
\label{eq:Pb}
\end{equation}
where $P_A$ to $P_E$ denote the respective parts of $A_1$ to $E$ proportional to $\exp(-2_0a)$.

First we will concentrate on the denominators of both expressions. The denominator of $P_L$ is

\begin{equation}
r_{D1} r_{D2}=\Big(\sin ^2(\theta ) ((k_0+q_{1e}) (k_\rho^2+q_{1o}^2) (k_0 \varepsilon_{1y}+q_{1o})-q_{1o} (k_0+q_{1o}) (k_0 q_{1e} \varepsilon_{1y}+q_{1o}^2))+
\label{eq:rD1rD2}
\end{equation}
\[q_{1o} (k_0+q_{1o}) (k_0 q_{1e} \varepsilon_{1y}+q_{1o}^2)\Big) \Big(\sin ^2(\theta +\varphi ) ((k_0+q_{2e}) (k_\rho^2+q_{2o}^2) (k_0 \varepsilon_{2y}+q_{2o})\]
\[-q_{2o} (k_0+q_{2o}) (k_0 q_{2e} \varepsilon_{2y}+q_{2o}^2))+q_{2o} (k_0+q_{2o}) (k_0 q_{2e} \varepsilon_{2y}+q_{2o}^2)\Big)\]
In order to compare this to Barash's expression, the numerator and denominator of the latter must be multiplied by $(q_{1o}^2+k_\rho^2\sin^2\theta)(q_{2o}^2+k_\rho^2\sin^2(\theta+\varphi))$. The denominator of $P_b$ is

\[\gamma(q_{1o}^2+k_\rho^2\sin^2\theta)(q_{2o}^2+k_\rho^2\sin^2(\theta+\varphi))=\]
\[\tfrac{1}{4} (k_0+q_{2o}) (k_0+q_{1o}) (k_\rho^2 \cos (2 \theta ) (q_{1o} (\varepsilon_{1y}-1)-\varepsilon_{1y} (k_0+q_{1e}))+\]
\[k_\rho^2 (\varepsilon_{1y} (k_0+q_{1e})-\varepsilon_{1y}q_{1o}+q_{1o})+2 (k_0 q_{1e} q_{1o} \varepsilon_{1y}+q_{1o}^3)) (k_\rho^2 \cos (2 (\theta +\varphi )) (q_{2o} (\varepsilon_{2y}-1)-\varepsilon_{2y} (k_0+q_{2e}))\]
\[+k_\rho^2 (\varepsilon_{2y} (k_0+q_{2e})-\varepsilon_{2y}q_{2o}+q_{2o})+2 (k_0 q_{2e} q_{2o} \varepsilon_{2y}+q_{2o}^3)),\]
which is identical to \eqref{eq:rD1rD2} due to the eigenvalues

\begin{equation}
q_{1o}^2+k_\rho^2=\varepsilon_{1y} (k_0^2+k_\rho^2)
\label{eq:qoq2o}
\end{equation} 
\[q_{2o}^2+k_\rho^2=\varepsilon_{2y} (k_0^2+k_\rho^2)\]

The numerator of \cref{eq:Pb} is far more complicated than its denominator. However it is considerably simplified by the assumptions $q_{1e}=q_{1o}$ and $q_{2e}=q_{2o}$. First we will prove that the numerator of \cref{eq:PL} is identical to that of \cref{eq:Pb} multiplied by $(q_{1o}^2+k_\rho^2\sin^2\theta)(q_{2o}^2+k_\rho^2\sin^2(\theta+\varphi))$ under these conditions. Later, these assumptions will be relaxed. If $q_{1e}=q_{1o}$ and $q_{2e}=q_{2o}$, the numerator of \cref{eq:Pb} will simplify to

\begin{equation}
p_{0b}=(q_{1o}^2+k_\rho^2\sin^2\theta)(q_{2o}^2+k_\rho^2\sin^2(\theta+\varphi))P_A|_{q_{1e}=q_{1o}, q_{2e}=q_{2o}}=
\label{eq:qe=qoq2o=q2e}
\end{equation}
\[-\Big((q_{2o}-k_0) (q_{1o}-k_0) (k_0 \varepsilon_{2y}+q_{2o}) (k_0 \varepsilon_{1y}+q_{1o})-(k_0+q_{2o}) (k_0+q_{1o}) (q_{2o}-k_0 \varepsilon_{2y}) (q_{1o}-k_0 \varepsilon_{1y})\Big)\times\]
\[(q_{1o}^2+k_\rho^2\sin^2\theta)(q_{2o}^2+k_\rho^2\sin^2(\theta+\varphi))\]
In this case the reflection matrices are both diagonal, hence the numerator of \cref{eq:PL} becomes
\[p_{7L}=-r_{ppN1} r_{ppN2}-r_{ssN1} r_{ssN2}|_{q_{1e}=q_{1o},q_{2o}=q_{2e}}=-2 (k_\rho^2 \sin ^2(\theta )+q_{1o}^2)(k_\rho^2 \sin ^2(\theta +\varphi )+q_{2o}^2)\times\]
\[ (k_0^4 \varepsilon_{2y} \varepsilon_{1y}-k_0^2 (q_{2o}^2 \varepsilon_{1y}+q_{2o} q_{1o} (\varepsilon_{2y}+\varepsilon_{1y}-\varepsilon_{1y}\varepsilon_{2y}-1)+q_{1o}^2 \varepsilon_{2y})+q_{2o}^2 q_{1o}^2) ,\]
which is equal to \cref{eq:qe=qoq2o=q2e}.

Now let $q_{1e}\neq q_{1o}$, while $q_{2o}=q_{2e}$. Now the numerator of $P_b$ is

\[(q_{1o}^2+k_\rho^2\sin^2\theta)(q_{2o}^2+k_\rho^2\sin^2(\theta+\varphi))P_A.\]
In this case only the term proportional to $(q_{1e}-q_{1o})$ needs to be considered, since we have already shown that the other terms are identical in the previous paragraph. Hence we are left with

\begin{equation}
-(q_{2o}^2+k_\rho^2\sin^2(\theta+\varphi))\times
\label{eq:(qe-qo)}
\end{equation}
\[2 (\varepsilon_{2y}-1) \varepsilon_{1y} (q_{1e}-q_{1o}) (k_\rho^2 \sin ^2(\theta ) (k_0^2 q_{2o}+k_\rho^2 q_{1o})- k_0^2 q_{1o} (-2 k_\rho^2 \sin ^2(\theta )+k_\rho^2-q_{2o} q_{1o}))\]

 Since $q_{2o}=q_{2e}$, $r_{spN2}=0=r_{psN2}$. Hence for $P_L$ the relevant term is

\[(-r_{ppN1} r_{ppN2}-r_{ssN1} r_{ssN2})|_{q_{2e}=q_{2o}}=\]
\[2 (k_\rho^2 \sin ^2(\theta )+q_{1o}^2) (k_0^4 \varepsilon_{2y} \varepsilon_{1y}-k_0^2 (q_{2o}^2 \varepsilon_{1y}+q_{2o} q_{1o} (\varepsilon_{2y}+\varepsilon_{1y}-\varepsilon_{1y}\varepsilon_{2y}-1)+q_{1o}^2 \varepsilon_{2y})+q_{2o}^2 q_{1o}^2) (k_\rho^2 \sin ^2(\theta +\varphi )+q_{2o}^2),\]
which can be simplified to

\[2 (q_{1e}-q_{1o}) (k_\rho^2 \sin ^2(\theta +\varphi )+q_{2o}^2) (\sin ^2(\theta ) (k_\rho^2 (-k_0^2 q_{2o} (\varepsilon_{2y}-1) \varepsilon_{1y}+k_0^2 q_{1o} \varepsilon_{2y}-q_{2o}^2 q_{1o})\]
\[+q_{1o} (k_0^2 \varepsilon_{2y}-q_{2o}^2) (k_0^2 \varepsilon_{1y}+q_{1o}^2))+k_0^2 q_{1o} \varepsilon_{1y} (-\varepsilon_{2y}k_0^2+q_{2o}^2+q_{2o} (q_{1o}-q_{1o} \varepsilon_{2y}))).\]
This expression is identical to \cref{eq:(qe-qo)} if \cref{eq:qoq2o} is taken into account.

Now let us make the reverse assumption, namely that $q_{1e}=q_{1o}$, but $q_{2e}\neq q_{2o}$. By the same token as before, we will focus on the term proportional to $q_{2e}-q_{2o}$, $p_2(q_{2e}-q_{2o})$. Because of the symmetry relation \cref{eq:p1p2} it follows that Barash's and Lifshitz-Lekner's versions of this expression should be identical as well. Nonetheless we will check this here:
\begin{equation}
i (q_{2e}- q_{2o})(q_{1o}^2+k_\rho^2\sin^2\theta)\varepsilon_{2y}(k_\rho^2 P_B \sin ^2(\theta +\varphi )+P_C)|_{q_{1e}=q_{1o}}=
\label{eq:dummy1}
\end{equation}
\[2 \varepsilon_{2y} (\varepsilon_{1y}-1) (q_{2e}-q_{2o}) (-k_\rho^2 \sin ^2(\theta )-q_{1o}^2) (k_\rho^2 \sin ^2(\theta +\varphi ) (k_0^2 (2 q_{2o}+q_{1o})+k_\rho^2 q_{2o})+k_0^2 q_{2o} (q_{2o} q_{1o}-k_\rho^2))\]
Here $P_E$ is omitted since it is proportional to $(q_{1e}-q_{1o})$. Now for $P_L$ we must evaluate

\[(-r_{ppN1} r_{ppN2}-r_{ssN1} r_{ssN2})|_{q_{1e}=q_{1o}}-(-r_{ppN1} r_{ppN2}-r_{ssN1} r_{ssN2})|_{q_{1e}=q_{1o},q_{2o}=q_{2e}}=\]
\[2 (q_{2e}-q_{2o}) (k_\rho^2 \sin ^2(\theta )+q_{1o}^2) (\sin ^2(\theta +\varphi ) (k_\rho^2 (q_{2o} (k_0^2 \varepsilon_{1y}-q_{1o}^2)-k_0^2 q_{1o} \varepsilon_{2y} (\varepsilon_{1y}-1))\]
\[+q_{2o} (k_0^2 \varepsilon_{2y}+q_{2o}^2) (k_0^2 \varepsilon_{1y}-q_{1o}^2))+k_0^2 q_{2o} \varepsilon_{2y} (-\varepsilon_{1y}k_0^2 +q_{1o} (q_{2o}-q_{2o} \varepsilon_{1y})+q_{1o}^2))\]
which is identical to \cref{eq:dummy1} under the conditions of \cref{eq:qoq2o}. 
Note that the third factor in the second term of Eq. (24) of Ref. \cite{Barash1978} should be $(r^2+\rho_1\rho_2)$, i.e. with a relative plus sign rather than a minus sign. This has been confirmed by Ref. \cite{Munday2005}, and now also here.

Finally, what remains is the term proportional to the product of the differences between the eigenvalues, $(q_{1e}-q_{1o})(q_{2e}-q_{2o})$. According to Barash, this is
\begin{equation}
\varepsilon_{2y} (i q_{2e}-i q_{2o}) (k_\rho^2 \sin ^2(\theta )+q_{1o}^2) (-P_E (2 k_\rho^2 \sin (\theta ) \cos (\varphi ) \sin (\theta +\varphi )
\label{eq:dummy3}
\end{equation}
\[-k_0^2 \sin ^2(\varphi ))+k_\rho^2 P_B \sin ^2(\theta +\varphi )+P_C)|_{\sim(q_{1e}-q_{1o})(q_{2e}-q_{2o})}=\]
\[-2 \varepsilon_{2y} \varepsilon_{1y} (q_{2e}-q_{2o}) (q_{1e}-q_{1o}) \Big(k_0^2 q_{1o} (q_{2o} (k_0^2+q_{2o} q_{1o})-2 k_0^2 q_{2o} \sin ^2(\varphi )-k_\rho^2 (q_{2o}-q_{1o}) \sin ^2(\theta +\varphi ))\]
\[+4 k_0^2 k_\rho^2 q_{2o} q_{1o} \sin (\theta ) \cos (\varphi ) \sin (\theta +\varphi )+\sin ^2(\theta ) (k_\rho^4 \sin ^2(\theta +\varphi ) (k_0^2+q_{2o} q_{1o})+k_0^2 k_\rho^2 q_{2o} (q_{2o}-q_{1o}))\Big)\]
The Lifshitz-Lekner equivalent of this expression is

\[-(r_{ppN1} r_{ppN2} +r_{psN1} r_{spN2}+r_{psN2} r_{spN1} +r_{ssN1} r_{ssN2})-\]
\[(-r_{ppN1} r_{ppN2}-r_{ssN1} r_{ssN2})|_{q_{1e}=q_{1o},q_{2o}=q_{2e}}-\]
\[(-r_{ppN1} r_{ppN2}-r_{ssN1} r_{ssN2})|_{q_{1e}=q_{1o}}-(-r_{ppN1} r_{ppN2}-r_{ssN1} r_{ssN2})|_{q_{2e}=q_{2o}}=\]
\[-2 (q_{2e}-q_{2o}) (q_{1e}-q_{1o}) \Big(k_0^2 q_{1o} \varepsilon_{1y} (q_{2o} \varepsilon_{2y} (\sin (2 \theta ) (k_0^2+k_\rho^2) \sin (2 (\theta +\varphi ))+k_0^2+q_{2o} q_{1o})\]
\[-\sin ^2(\theta +\varphi ) (k_0^2 q_{2o} \varepsilon_{2y}+k_\rho^2 (q_{2o}-q_{1o} \varepsilon_{2y})+q_{2o}^3))+\sin ^2(\theta ) (\sin ^2(\theta +\varphi ) (k_\rho^4 (k_0^2 \varepsilon_{2y} \varepsilon_{1y}+q_{2o} q_{1o})\]
\[+k_\rho^2 q_{2o} q_{1o} (k_0^2 (\varepsilon_{2y}+\varepsilon_{1y})+q_{2o}^2+q_{1o}^2)+q_{2o} q_{1o} (k_0^2 \varepsilon_{2y}+q_{2o}^2) (k_0^2 \varepsilon_{1y}+q_{1o}^2))-k_0^2 q_{2o} \varepsilon_{2y} (k_0^2 q_{1o} \varepsilon_{1y}+k_\rho^2 (q_{1o}-q_{2o} \varepsilon_{1y})+q_{1o}^3))\Big)\]
which is identical to \cref{eq:dummy3} if \cref{eq:qoq2o} is considered. This completes the proof that 

\begin{equation}
\boxed{P_b=P_L}
\label{eq:Pb=PL}
\end{equation}

\subsection{The Coefficient $Q$}\label{sec:CoeffQ}

\noindent Finally we direct our attention towards the terms proportional to $\exp(-4k_0 a)$, denoted by $Q_L$:
\begin{equation}
Q_L=\frac{1}{r_{D1}^2 r_{D2}^2}\Big(r_{ppN1} r_{ppN2} r_{ssN1} r_{ssN2}-r_{ppN1} r_{psN2} r_{spN2} r_{ssN1}
\label{eq:QL}
\end{equation}
\[-r_{ppN2} r_{psN1} r_{spN1} r_{ssN2} +r_{psN1} r_{psN2} r_{spN1} r_{spN2}\Big)\]

In analogy to the previous subsection, the same coefficient according to Barash is
\[Q_b=\frac{Q_A}{\gamma}+\]
\begin{equation}
\frac{i\varepsilon_{2y} (q_{2e}-q_{2o}) (k_\rho^2 Q_B \sin ^2(\theta +\varphi )+Q_C)}{\gamma(k_\rho^2 \sin ^2(\theta +\varphi )+q_{2o}^2)}
\label{eq:Qb}
\end{equation}
where $Q_A$ to $Q_C$ denote the respective parts of $A_1$ to $C$ proportional to $\exp(-4 k_0 a)$. $Q_E$ is omitted since $E$ is proportional to $\exp(-2k_0 a)$ only.

The denominator of $Q_L$ is actually the square of that of $P_L$:

\[r_{D1}^2 r_{D2}^2=\gamma^2(q_{1o}^2+k_\rho^2\sin^2\theta)^2(q_{2o}^2+k_\rho^2\sin^2(\theta+\varphi))^2\]
Hence in order to compare $Q_L$ to $Q_b$, the numerator of the latter must be multiplied by a factor of $\gamma(q_{1o}^2+k_\rho^2\sin^2\theta)^2(q_{2o}^2+k_\rho^2\sin^2(\theta+\varphi))^2$. After all, the denominator of $Q_b$ is simply $\gamma$.

As before, we will start with the simplest case: $q_{1e}=q_{1o}, q_{2e}=q_{2o}$. The numerator of $Q_L$ will simplify to

\[r_{ppN1} r_{ppN2} r_{ssN1} r_{ssN2}|_{q_{1e}=q_{1o},q_{2o}=q_{2e}}=\]
\[(q_{2o}^2-k_0^2) (q_{1o}^2-k_0^2) (q_{2o}^2-k_0^2 \varepsilon_{2y}^2) (q_{1o}^2-k_0^2 \varepsilon_{1y}^2) (k_\rho^2 \sin ^2(\theta )+q_{1o}^2)^2 (k_\rho^2 \sin ^2(\theta +\varphi )+q_{2o}^2)^2,\]
of which Barash's equivalent is

\[Q_A\gamma(q_{1o}^2+k_\rho^2\sin^2\theta)^2(q_{2o}^2+k_\rho^2\sin^2(\theta+\varphi))^2|_{q_{1e}=q_{1o},q_{2o}=q_{2e}}=\]
\[(q_{2o}^2-k_0^2) (q_{1o}^2-k_0^2) (q_{2o}^2-k_0^2 \varepsilon_{2y}^2) (q_{1o}^2-k_0^2 \varepsilon_{1y}^2) (k_\rho^2 \sin ^2(\theta )+q_{1o}^2)^2 (k_\rho^2 \sin ^2(\theta +\varphi )+q_{2o}^2)^2,\]
which is clearly identical to the Lifshitz-Lekner expression.

Next we will assume again that $q_{2e}=q_{2o}$ and that $q_{1e}\neq q_{1o}$. However, contrary to the previous subsection, now the numerator of $Q$ is a quadratic function of $q_{1e}-q_{1o}$. In the previous paragraph, it has been shown that the constant term in this case is identical according to both Lifshitz-Lekner and Barash. The latter expression in this case is

\[Q_A|_{q_{2e}=q_{2o}}=(q_{2o}-k_0)(q_{1o}-k_0)(q_{2o}-\varepsilon_{2y}k_0)(q_{1o}-\varepsilon_{1y}k_0)+\]
\[\frac{(q_{1e}-q_{1o})(q_{2o}-k_0)(q_{1o}-k_0)(q_{2o}-\varepsilon_{2y}k_0)\varepsilon_{1y}(k_\rho^2\sin^2\theta-k_0q_{1o})}{q_{1o}^2+k_\rho\sin^2\theta}\]

which must be be multiplied by

\[\gamma|_{q_{2e}=q_{2o}} (k_\rho^2\sin^2\theta+q_{1o}^2)^2(q_{2o}^2+\sin^2(\theta+\varphi))^2=\]
\[(q_{2o}^2+k_\rho^2\sin(\theta+\varphi))^2(k_0+q_{2o})(k_0+q_{1o})(q_{2o}+\varepsilon_{2y}k_0)(q_{1o}^2+k_\rho^2\sin^2\theta)\times\]
\[\Big[(q_{1o}+\varepsilon_{1y}k_0)(q_{1o}^2+k_\rho^2\sin^2\theta)+(q_{1e}-q_{1o})\varepsilon_{1y}(k_0q_{1o}+k_\rho^2\sin^2\theta)\Big].\]
Here we are concerned only with the terms proportional to $(q_{1e}-q_{1o})$ and $(q_{1e}-q_{1o})^2$ of the product between these expressions. The former is according to Barash:

\[q_{1b}=\varepsilon_{1y} (k_0-q_{2o}) (k_0+q_{2o}) (k_0-q_{1o}) (k_0+q_{1o}) (q_{1e}-q_{1o}) (q_{2o}-k_0 \varepsilon_{2y}) (k_0 \varepsilon_{2y}+q_{2o}) (k_\rho^2 \sin ^2(\theta )+q_{1o}^2) ((k_0 \varepsilon_{1y}+q_{1o})\times\]
\[ (k_0 q_{1o}-k_\rho^2 \sin ^2(\theta ))-(q_{1o}-k_0 \varepsilon_{1y}) (k_0 q_{1o}+k_\rho^2 \sin ^2(\theta )) (k_\rho^2 \sin ^2(\theta +\varphi )+q_{2o}^2)^2),\]
which can be simplified to

\begin{equation}
q_{1b}=-\tfrac{1}{8} q_{1o} \varepsilon_{1y} (k_0-q_{2o}) (k_0+q_{2o}) (k_0-q_{1o}) (k_0+q_{1o}) (k_0 \varepsilon_{2y}-q_{2o}) (k_0 \varepsilon_{2y}+q_{2o})\times
\label{eq:dummy7}
\end{equation}
\[ ( -k_\rho^2\cos (2 \theta )+k_\rho^2+2 q_{1o}^2) (2 k_0^2 \varepsilon_{1y}+k_\rho^2 \cos (2 \theta )-k_\rho^2) ( -k_\rho^2\cos (2 (\theta +\varphi ))+k_\rho^2+2 q_{2o}^2)^2,\]
and the latter is
\begin{equation}
q_{4b}=\varepsilon_{1y}^2 (q_{2o}^2-k_0^2) (q_{1o}^2-k_0^2) (q_{1e}-q_{1o})^2 (q_{2o}^2-k_0^2 \varepsilon_{2y}^2) (k_\rho^2 \sin ^2(\theta )-k_0 q_{1o}) (k_0 q_{1o}+k_\rho^2 \sin ^2(\theta )) (k_\rho^2 \sin ^2(\theta +\varphi )+q_{2o}^2)^2
\label{eq:dummy6}
\end{equation}
\[=(q_{2o}^2-k_0^2) (q_{1o}^2-k_0^2) (q_{2o}^2-k_0^2 \varepsilon_{2y}^2) (q_{1o}^2-k_0^2 \varepsilon_{1y}^2) (k_\rho^2 \sin ^2(\theta )+q_{1o}^2)^2 (k_\rho^2 \sin ^2(\theta +\varphi )+q_{2o}^2)^2.\]

The equivalent expressions according to Lifsitz-Lekner can be obtained by simplifying the factors $r_{ppN1} r_{ppN2}$, $r_{ssN1} r_{ssN2}$, $r_{spN1} r_{ssN2}$, and $-r_{ppN2} r_{psN1}$ first. From this we gather the terms proportional to $q_{1e}-q_{1o}$:

\[q_{1L}=(k_0-q_{2o}) (k_0+q_{2o}) (k_0+q_{1o}) (q_{1o}-q_{1e}) (k_0 \varepsilon_{2y}+q_{2o}) (q_{2o}-k_0 \varepsilon_{2y}) (q_{1o}-k_0 \varepsilon_{1y}) (k_\rho^2 \sin ^2(\theta )+q_{1o}^2) \times\]
\[ (k_\rho^2 \sin ^2(\theta +\varphi )+q_{2o}^2)^2(\sin ^2(\theta ) (k_0^2 q_{1o} \varepsilon_{1y}+k_\rho^2 (k_0 \varepsilon_{1y}+q_{1o})+q_{1o}^3)+k_0 q_{1o} \varepsilon_{1y} (q_{1o}-k_0))+\]
\[(k_0-q_{2o}) (k_0+q_{2o}) (k_0-q_{1o}) (q_{1e}-q_{1o}) (k_0 \varepsilon_{2y}+q_{2o}) (q_{2o}-k_0 \varepsilon_{2y}) (k_0 \varepsilon_{1y}+q_{1o}) (k_\rho^2 \sin ^2(\theta )+q_{1o}^2) \times\]
\[ (k_\rho^2 \sin ^2(\theta +\varphi )+q_{2o}^2)^2(\sin ^2(\theta ) (k_0^2 q_{1o} \varepsilon_{1y}+k_\rho^2 (q_{1o}-k_0 \varepsilon_{1y})+q_{1o}^3)-k_0 q_{1o} \varepsilon_{1y} (k_0+q_{1o})),\]
which is identical to \cref{eq:dummy7} with the eigenvalues  \cref{eq:qoq2o}. So

\begin{equation}
q_{1b}=q_{1L}
\label{eq:q1bq1L}
\end{equation}

The terms proportional to $(q_{1e}-q_{1o})^2$ are

\[q_{4L}=-(k_0-q_{2o}) (k_0+q_{2o}) (q_{1e}-q_{1o})^2 (k_0 \varepsilon_{2y}+q_{2o}) (q_{2o}-k_0 \varepsilon_{2y})\times\]
\[ (k_\rho^2 \sin ^2(\theta +\varphi )+q_{2o}^2)^2 (\sin ^2(\theta ) (k_0^2 q_{1o} \varepsilon_{1y}+k_\rho^2 (k_0 \varepsilon_{1y}+q_{1o})+q_{1o}^3)+k_0 q_{1o} \varepsilon_{1y} (q_{1o}-k_0))\times\]
\[ (\sin ^2(\theta ) (k_0^2 q_{1o} \varepsilon_{1y}+k_\rho^2 (q_{1o}-k_0 \varepsilon_{1y})+q_{1o}^3)-k_0 q_{1o} \varepsilon_{1y} (k_0+q_{1o}))+\]
\[k_0^2 q_{1o}^2 \varepsilon_{1y}^2 \sin ^2(2 \theta ) (k_0^2+k_\rho^2) (q_{2o}-k_0) (k_0+q_{2o}) (q_{1e}-q_{1o})^2 (k_0 \varepsilon_{2y}+q_{2o}) (q_{2o}-k_0 \varepsilon_{2y}) (k_\rho^2 \sin ^2(\theta +\varphi )+q_{2o}^2)^2,\]
which is identical to \cref{eq:dummy6} with the eigenvalues  \cref{eq:qoq2o}. Hence

\begin{equation}
q_{4b}=q_{4L}
\label{eq:q4bq4L}
\end{equation}

We will now assume again that $q_{1e}= q_{1o}$ and that  $q_{2e}\neq q_{2o}$, and focus on the terms proportional to $q_{2e}-q_{2o}$ and $(q_{2e}-q_{2o})^2$. The term proportional to $(q_{2e}-q_{2o})^2$, according to Barash, can be simplified to

\begin{equation}
q_{5b}=-\tfrac{1}{16} \varepsilon_{2y}^2 (q_{2e}-q_{2o})^2(k_0-q_{2o}) (k_0+q_{2o}) (k_0-q_{1o}) (k_0+q_{1o}) (k_0 \varepsilon_{1y}-q_{1o}) (k_0 \varepsilon_{1y}+q_{1o})\times
\label{eq:dummy10}
\end{equation}
\[ (-k_\rho^2 \cos (2 \theta )+k_\rho^2+2 q_{1o}^2)^2 (2 k_0 q_{2o}-k_\rho^2 \cos (2 (\theta +\varphi ))+k_\rho^2) (2 k_0 q_{2o}+k_\rho^2 \cos (2 (\theta +\varphi ))-k_\rho^2)\]
According to Lekner-Lifshitz this term is

\[q_{5L}=(q_{1o}^2-k_0^2) (q_{2e}-q_{2o})^2 (k_0 \varepsilon_{1y}+q_{1o}) (q_{1o}-k_0 \varepsilon_{1y}) (k_\rho^2 \sin ^2(\theta )+q_{1o}^2)^2\times\]
\[ (\sin ^2(\theta +\varphi ) (k_0^2 q_{2o} \varepsilon_{2y}+k_\rho^2 (k_0 \varepsilon_{2y}+q_{2o})+q_{2o}^3)+k_0 q_{2o} \varepsilon_{2y} (q_{2o}-k_0)) (\sin ^2(\theta +\varphi )\times\]
\[ (k_0^2 q_{2o} \varepsilon_{2y}+k_\rho^2 (q_{2o}-k_0 \varepsilon_{2y})+q_{2o}^3)-k_0 q_{2o} \varepsilon_{2y} (k_0+q_{2o}))+\]
\[ k_0^2 q_{2o}^2 \varepsilon_{2y}^2 (k_0^2+k_\rho^2) (k_0-q_{1o}) (k_0+q_{1o}) (q_{2e}-q_{2o})^2 \sin ^2(2 (\theta +\varphi ))(k_0 \varepsilon_{1y}-q_{1o}) (k_0 \varepsilon_{1y}+q_{1o}) (k_\rho^2 \sin ^2(\theta )+q_{1o}^2)^2,\]
which is the same as \cref{eq:dummy10} with the eigenvalue relations \cref{eq:qoq2o}. In other words

\begin{equation}
q_{5b}=q_{5L}
\label{eq:q5bq5L}
\end{equation}

The symmetry relation \cref{eq:q1q2} shows that the Lekner-Lifshitz version of the term proportional to $q_{2o}-q_{2e}$ is identical to its Barash's version:

\begin{equation}
q_{2b}=q_{2L}
\label{eq:q2bq2L}
\end{equation}

Finally, we arrive at the mixed terms, i.e. those proportional to $(q_{1e}-q_{1o})^j(q_{2o}-q_{2e})^k$ with $j,k\in\{1,2\}$. There are in total four such terms, three of which are independent. First we will establish the Barash (denoted by subscript $b$ and Lifshitz-Lekner (subscript $L$) variants of these coefficients independently. 

Let us start with the Lifshitz variant. This has the advantage that each coefficient can be written as a product of terms associated with medium 1 and those associated with medium 2. This makes it possible to calculate only the terms associated with the first medium, and then multiply that with a similar expression corresponding to the other medium. The part of $Q_L$  associated with medium 1, denoted by $Q_{1L}$, can be written as follows:

\[Q_{1L}=
{q_{1o}}^2 ({k_0}^2 {\varepsilon_{1y}}^2 \sin ^2(2 \theta ) ({k_0}^2+{k_\rho}^2) ({q_{1e}}-{q_{1o}})^2+({k_0}^2-{q_{1o}}^2) ({k_0}^2 {q_{1e}}^2 {\varepsilon_{1y}}^2-{q_{1o}}^4))\]
\[-2 {q_{1o}} \sin ^2(\theta ) (-{k_0}^4 {q_{1e}} {\varepsilon_{1y}}^2 ({k_\rho}^2+{q_{1o}} ({q_{1o}}-{q_{1e}}))\]
\[+{k_0}^2 ({k_\rho}^2 {q_{1o}} ({q_{1e}}^2 {\varepsilon_{1y}} ({\varepsilon_{1y}}+1)-2 {q_{1e}} {q_{1o}} {\varepsilon_{1y}}+{q_{1o}}^2 ({\varepsilon_{1y}}+1))+{q_{1o}}^3 {\varepsilon_{1y}} ({q_{1e}}-{q_{1o}})^2)-{k_\rho}^2 {q_{1e}} {q_{1o}}^4+{q_{1o}}^6 ({q_{1o}}-{q_{1e}}))\]
\[+\sin ^4(\theta ) ({k_0}^4 {\varepsilon_{1y}}^2 ({k_\rho}^2+{q_{1o}} ({q_{1o}}-{q_{1e}}))^2-{k_0}^2 ({k_\rho}^4 ({q_{1e}}^2 {\varepsilon_{1y}}^2+{q_{1o}}^2)-2 {k_\rho}^2 {q_{1o}}^2 {\varepsilon_{1y}} ({q_{1e}}-{q_{1o}})^2-2 {q_{1o}}^4 {\varepsilon_{1y}} ({q_{1e}}-{q_{1o}})^2)\]
\[+({k_\rho}^2 {q_{1e}} {q_{1o}}+{q_{1o}}^3 ({q_{1e}}-{q_{1o}}))^2)
\]
which has to multiplied with the same expression transformed to medium 2 to obtain the total $Q_L$. From now on we will limit ourselves to the part of $Q_L$ corresponding to medium 1 only. The goal is now to write this expression as a second order polynomial in $q_{1e}-q_{1o}$. Then the total $Q_L$ is a product of this polynomial with the corresponding polynomial for medium 2. For this, we gather the coefficients $q_{0L}$ to $q_{8L}$.  The first step is to expand $Q_{1L}$ completely as a polynomial in $q_{1e}$. This leads to the rather lengthy expression, so it is more insightful to write this implicitly in terms of the dummy variable coefficients $q'_{1i}$, (where the first subscript labels the medium, and the second one the order) i.e.

\[Q_{1L}=q_{10}'+q_{11}'q_{1e}+q_{12}'q_{1e}^2\]
which can be transformed into a polynomial in $q_{1e}-q_{1o}$ as

\[Q_{1L}=q_{10}+q_{11}(q_{1e}-q_{1o})+q_{12}(q_{1e}-q_{1o})^2\]
which leads to the following relations between the dummy coefficients:

\[q_{12}=q_{12}'\]
\[q_{11}=q_{11}'+2q_{12} q_{1o}\]
\[q_{10}=q_{10}'+q_{11}q_{1o}-q_{12}q_{1o}^2.\]
Now the same procedure is repeated for medium 2, leading to the analogous coefficients $q_{20}$, $q_{21}$, and $q_{22}$. Then the total $Q_L$ is written as

\[Q_L=Q_{1L}Q_{2L}=(q_{10}+q_{11}(q_{1e}-q_{1o})+q_{12}(q_{1e}-q_{1o})^2)(q_{20}+q_{21}(q_{2e}-q_{2o})+q_{22}(q_{2e}-q_{2o})^2)\]
which leads to the following relevant Lifshitz coefficients, now written explicitly:

\begin{equation}
q_{3L}=q_{11}q_{21}=
\label{eq:q3L}
\end{equation}
\[\frac{1}{4} {q_{2o}} {q_{1o}} ({k_\rho}^2 (1-\cos (2 \theta ))+2 {q_{1o}}^2) ({k_\rho}^2 (1-\cos (2 (\theta +\varphi )))+2 {q_{2o}}^2)\times\]
\[ ({k_\rho}^2 ({k_0}^2 {\varepsilon_{1y}}^2-{q_{1o}}^2)- {\varepsilon_{1y}}^2{k_0}^4+2 {k_0}^2 {q_{1o}}^2 {\varepsilon_{1y}}^2+\cos (2 \theta ) ({k_\rho}^2 ({q_{1o}}^2-{k_0}^4 {\varepsilon_{1y}}^2)-{k_0}^2 {\varepsilon_{1y}}^2+{q_{1o}}^4)-{q_{1o}}^4)\times\]
\[ ({k_\rho}^2 ({k_0}^2 {\varepsilon_{2y}}^2-{q_{2o}}^2)+2 {k_0}^2 {q_{2o}}^2 {\varepsilon_{2y}}^2+\cos 2 (\theta +\varphi ) ({k_\rho}^2 ({q_{2o}}^2-{k_0}^2 {\varepsilon_{2y}}^2)-{\varepsilon_{2y}}^2{k_0}^4 +{q_{2o}}^4)-{q_{2o}}^4)\]

\begin{equation}
q_{7L}=q_{11}q_{22}=
\label{eq:q7L}
\end{equation}
\[-\frac{1}{2} {q_{1o}} ({k_\rho}^2 (1-\cos (2 \theta ))+2 {q_{1o}}^2)\times\]
\[ ({k_\rho}^2 ({k_0}^2 {\varepsilon_{1y}}^2-{q_{1o}}^2)-{k_0}^4 {\varepsilon_{1y}}^2+2 {k_0}^2 {q_{1o}}^2 {\varepsilon_{1y}}^2+\cos (2 \theta ) ({k_\rho}^2 ({q_{1o}}^2-{k_0}^2 {\varepsilon_{1y}}^2)-{k_0}^4 {\varepsilon_{1y}}^2+{q_{1o}}^4)-{q_{1o}}^4)\times\]
\[ ({k_0}^2 {q_{2o}}^2 {\varepsilon_{2y}}^2 (({k_0}^2+{k_\rho}^2) \sin ^2(2 (\theta +\varphi ))+{k_0}^2-{q_{2o}}^2)-2 {k_0}^2 {q_{2o}}^2 {\varepsilon_{2y}} \sin ^2(\theta +\varphi ) ({k_0}^2 {\varepsilon_{2y}}+{k_\rho}^2 ({\varepsilon_{2y}}+1)+{q_{2o}}^2)\]
\[+\sin ^4(\theta +\varphi ) ({k_\rho}^4 ({q_{2o}}^2-{k_0}^2 {\varepsilon_{2y}}^2)+2 {k_\rho}^2 ({k_0}^2 {q_{2o}}^2 {\varepsilon_{2y}}+{q_{2o}}^4)+({k_0}^2 {q_{2o}} {\varepsilon_{2y}}+{q_{2o}}^3)^2))\]
The reason why we consider $q_{7L}$ and not $q_{6L}$ will become apparent later.
\begin{equation}
q_{8L}=q_{12}q_{22}=
\label{eq:q8L}
\end{equation}
\[({k_0}^2 {q_{1o}}^2 {\varepsilon_{1y}}^2 (\sin ^2(2 \theta ) ({k_0}^2+{k_\rho}^2)+{k_0}^2-{q_{1o}}^2)-2 {k_0}^2 {q_{1o}}^2 {\varepsilon_{1y}} \sin ^2(\theta ) ({k_0}^2 {\varepsilon_{1y}}+{k_\rho}^2 ({\varepsilon_{1y}}+1)+{q_{1o}}^2)\]
\[+\sin ^4(\theta ) ({k_\rho}^4 ({q_{1o}}^2-{k_0}^2 {\varepsilon_{1y}}^2)+2 {k_\rho}^2 ({k_0}^2 {q_{1o}}^2 {\varepsilon_{1y}}+{q_{1o}}^4)+({k_0}^2 {q_{1o}} {\varepsilon_{1y}}+{q_{1o}}^3)^2))\times\]
\[ ({k_0}^2 {q_{2o}}^2 {\varepsilon_{2y}}^2 (({k_0}^2+{k_\rho}^2) \sin ^2(2 (\theta +\varphi ))+{k_0}^2-{q_{2o}}^2)-2 {k_0}^2 {q_{2o}}^2 {\varepsilon_{2y}} \sin ^2(\theta +\varphi ) ({k_0}^2 {\varepsilon_{2y}}+{k_\rho}^2 ({\varepsilon_{2y}}+1)+{q_{2o}}^2)\]
\[+\sin ^4(\theta +\varphi ) ({k_\rho}^4 ({q_{2o}}^2-{k_0}^2 {\varepsilon_{2y}}^2)+2 {k_\rho}^2 ({k_0}^2 {q_{2o}}^2 {\varepsilon_{2y}}+{q_{2o}}^4)+({k_0}^2 {q_{2o}} {\varepsilon_{2y}}+{q_{2o}}^3)^2))\]

Next we will move on to the Barash variant of these coefficients, i.e. $q_{3b}$, $q_{7b}$, and $q_{8b}$. Unfortunately these are not so straightforwardly determined, since Barash's result is not separated into a product of terms associated with medium 1 and medium 2. Rather, $Q_b=Q_{1b}+Q_{2b}$ is split into two terms as in \cref{eq:Qb}, where both terms contain a mixture of expressions depending on both media. The first term of \cref{eq:Qb} is a quadratic polynomial in $q_{1e}-q_{1o}$ and $q_{2e}-q_{2o}$:

\[Q_{1b}\equiv Q_A\gamma(q_{1o}^2+k_\rho^2\sin^2\theta)^2(q_{2o}^2+k_\rho^2\sin^2(\theta+\varphi))^2=\]
\[d_0+d_1 (q_{1e}-q_{1o})+d_2(q_{2e}-q_{2o})+d_3(q_{1e}-q_{1o})(q_{2e}-q_{2o})+d_4(q_{1e}-q_{1o})^2+d_5(q_{1e}-q_{1o})^2 (q_{2e}-q_{2o})\]
for some dummy coefficients $d_i$. So this term contributes only to $q_{3b}$ and $q_{6b}$, and the other mixed terms must be extracted from the second term

\[Q_{2b}=(i\varepsilon_{2y} (q_{2e}-q_{2o}) (k_\rho^2 Q_B \sin ^2(\theta +\varphi )+Q_C)\gamma(q_{1o}^2+k_\rho^2\sin^2\theta)^2(q_{2o}^2+k_\rho^2\sin^2(\theta+\varphi))=\]
\[d_5(q_{2e}-q_{2o})+d_6(q_{2e}-q_{2o})^2+d_7(q_{1e}-q_{1o})(q_{2e}-q_{2o})+d_8(q_{1e}-q_{1o})^2(q_{2e}-q_{2o})+\]
\[q_{7b}(q_{1e}-q_{1o})(q_{2e}-q_{2o})^2+q_{8b}(q_{1e}-q_{1o})^2(q_{2e}-q_{2o})^2\]
Now it becomes apparent why we considered $q_{7L}$ in \cref{eq:q7L}: it is more convenient since $q_{7b}$ is contained in one term, unlike $q_{6b}$. The same applies to $q_{8b}$, which is the most convenient to start with, as it is the highest order term. Simply multiplying the terms $~(q_{1e}-q_{1o})$ and $~(q_{2e}-q_{2o})$ within the parentheses yields
\begin{equation}
q_{8b}= {\varepsilon_{2y}}^2 {\varepsilon_{1y}}^2 (i {k_0}+i {q_{2o}}) (i {k_0}+i {q_{1o}}) ({k_0} {q_{1o}}+{k_\rho}^2 \sin ^2(\theta )) \times
\label{eq:q8b}
\end{equation}
\[({k_0} {q_{2o}}+({k_\rho} \sin (\theta ) \cos (\varphi )+{k_\rho} \cos (\theta ) \sin (\varphi ))^2) ({k_\rho}^2 (i {q_{2o}}-i {k_0}) (i {q_{1o}}-i {k_0}) \sin ^2(\theta +\varphi ) ({k_\rho}^2 \sin ^2(\theta )-{k_0} {q_{1o}})\]
\[-{k_0} {q_{2o}}  (i {q_{2o}}-i {k_0}) (i {q_{1o}}-i {k_0}) ({k_\rho}^2 \sin ^2(\theta )-{k_0} {q_{1o}}))\]
which is identical to \cref{eq:q8L} under the condition of \cref{eq:qoq2o}. Hence

\begin{equation}
q_{8b}=q_{8L}
\label{eq:q8bq8L}
\end{equation}

Next we will direct our attention to the coefficient $q_{7b}$. This one is a bit harder to determine. It is useful to remember that $Q_{2b}$ is essentially a polynomial of the form 

\[Q_{2b}=a_0 y (a_1 +a_2 x) (a_3+a_4 y)(a_5+a_6x) \]
where $x=q_{1e}-q_{1o}$ and $y=q_{2e}-q_{2o}$ and the coefficients $a_i$ are complicated expressions. Here $q_{7b}$ would be the coefficient of this polynomial proportional to $xy^2$, which is

\[a_0 a_4(a_2 a_5+a_1 a_6).\]
More explicitly, in our case this coefficient is

\begin{equation}
q_{7b}=-\frac{1}{8} {q_{1o}} {\varepsilon_{2y}}^2 {\varepsilon_{1y}} ({k_0}-{q_{2o}}) ({k_0}+{q_{2o}}) ({k_0}-{q_{1o}}) ({k_0}+{q_{1o}}) ({k_\rho}^2 (1-\cos (2 \theta )+2 {q_{1o}}^2) (2 {k_0}^2 {\varepsilon_{1y}}+{k_\rho}^2 (\cos (2 \theta )-1))\times
\label{eq:q7b}
\end{equation}
\[ (2 {k_0} {q_{2o}}+{k_\rho}^2 (-\cos (2 (\theta +\varphi )))+{k_\rho}^2) (2 {k_0} {q_{2o}}+{k_\rho}^2 \cos (2 (\theta +\varphi ))-{k_\rho}^2),\]
which identical to \cref{eq:q7L} under the conditions of \cref{eq:qoq2o}. Hence 

\begin{equation}
q_{7b}=q_{7L}
\label{eq:q7bq7L}
\end{equation}

Due to the symmetry relation \cref{eq:q6q7} the same equality can be claimed for the sixth coefficients:
\begin{equation}
q_{6b}=q_{6L}.
\label{eq:q6bq6L}
\end{equation}
Finally, we consider the most difficult coefficient to obtain, $q_{3b}$. In accordance with the previous analogy, we can write the total $Q_b$ as

\[Q_b=a_0 y (a_1 +a_2 x) (a_3+a_4 y)(a_5+a_6x)+b_0(b_1+b_2 x)(b_3+b_4 y)(b_5+b_6 y)\]
Here $q_{3b}$ would be the coefficient multiplied with $xy$, which is

\[{a_0} {a_1} {a_3} {a_6}+{a_0} {a_2} {a_3} {a_5}+{b_0} {b_1} {b_4} {b_6}+{b_0} {b_2} {b_3} {b_6}\]
Explicitly $q_{3b}$ is then, after simplification,

\begin{equation}
q_{3b}=\frac{1}{4} {q_{2o}} {q_{1o}} {\varepsilon_{2y}} {\varepsilon_{1y}} ({k_0}-{q_{2o}}) ({k_0}+{q_{2o}}) ({k_0}-{q_{1o}}) ({k_0}+{q_{1o}}) ({k_\rho}^2 (1-\cos (2 \theta ))+2 {q_{1o}}^2)\times 
\label{eq:q3b}
\end{equation}
\[(2 {k_0}^2 {\varepsilon_{1y}}+{k_\rho}^2 \cos (2 \theta )-{k_\rho}^2) ({k_\rho}^2 (1-\cos (2 (\theta +\varphi )))+2 {q_{2o}}^2) (2 {k_0}^2 {\varepsilon_{2y}}+{k_\rho}^2 \cos (2 (\theta +\varphi ))-{k_\rho}^2),\]
which is identical to \cref{eq:q3L} under conditions of \cref{eq:qoq2o}. Hence

\begin{equation}
q_{3b}=q_{3L}
\label{eq:q3bq3L}.
\end{equation}

In this subsection we have taken advantage of the symmetry relations \cref{eq:q1q2} to obtain \cref{eq:q2bq2L} and \cref{eq:q6q7} to obtain \cref{eq:q6bq6L}. So the natural next step is to prove that \cref{eq:q1q2,eq:q6q7} hold indeed. This is simply a matter of determining the relevant coefficients, starting with 

\begin{equation}
q_{2b}=\frac{1}{8} {q_{2o}} {\varepsilon_{2y}} ({q_{2o}}^2-{k_0}^2) ({k_0}^2-{q_{1o}}^2) ({k_0}^2 {\varepsilon_{1y}}^2-{q_{1o}}^2)\times \label{eq:q2b}
\end{equation}
\[({k_\rho}^2 (1-\cos (2 \theta ))+2 {q_{1o}}^2)^2 ({k_\rho}^2 (1-\cos (2 (\theta +\varphi )))+2 {q_{2o}}^2) (2 {k_0}^2 {\varepsilon_{2y}}+{k_\rho}^2 \cos (2 (\theta +\varphi ))-{k_\rho}^2),\]
which is the transformed version of $q_{1b}$ from \cref{eq:dummy7}. Now finally we move on to the remaining coefficient

\begin{equation}
q_{6b}=-\frac{1}{8} {q_{2o}} {\varepsilon_{2y}} {\varepsilon_{1y}}^2 ({k_0}-{q_{2o}}) ({k_0}+{q_{2o}}) ({k_0}-{q_{1o}}) ({k_0}+{q_{1o}})\times 
\label{eq:q6b}
\end{equation}
\[(2 {k_0} {q_{1o}}+{k_\rho}^2 (1-\cos (2 \theta ))) (2 {k_0} {q_{1o}}+{k_\rho}^2 \cos (2 \theta )-{k_\rho}^2) ({k_\rho}^2 (1-\cos (2 (\theta +\varphi )))+2 {q_{2o}}^2)\times\]
\[ (2 {k_0}^2 {\varepsilon_{2y}}+{k_\rho}^2 (\cos (2 (\theta +\varphi ))-1)),\]
which is \cref{eq:q7b} with the media swapped.

This completes the proof that 

\begin{equation}
\boxed{Q_L=Q_b}
\label{eq:QLQb}
\end{equation}
Since $P_b=P_L$ (see \cref{sec:coeffP}) and the constant term of $D_b$ equals unity (see \cref{Barash1}), this also proves that the result {\cref{eq:Dn}} for $\mu_x=1=\mu_y$ is identical to Eq. (21) of Ref. \cite{Barash1978}. To our knowledge this is the first independent proof of this result.


\end{document}